\documentclass[aps,pre,twocolumn,superscriptaddress,amsmath,amssymb,showpacs]{revtex4-1}
\usepackage{amsmath}
\usepackage{amssymb}
\usepackage{graphicx}
\usepackage{float}
\usepackage{subfigure}
\usepackage{color}
\usepackage{hyperref}
\usepackage{comment}
\usepackage{makecell}

\begin{document}
\title{Dimensional crossover in surface growth on rectangular substrates}
\author{Ismael S. S. Carrasco}
\email{ismael.carrasco@unb.br}
\affiliation{Departamento de F\'isica, Universidade Federal de Vi\c cosa, 36570-900, Vi\c cosa, MG, Brazil}
\affiliation{International Center of Physics, Institute of Physics, University of Brasilia, 70910-900, Brasilia, Federal District, Brazil}
\author{Tiago J. Oliveira}
\email{tiago@ufv.br}
\affiliation{Departamento de F\'isica, Universidade Federal de Vi\c cosa, 36570-900, Vi\c cosa, MG, Brazil}
\date{\today}

\begin{abstract}
In a recent work [Phys. Rev. E {\bf 109}, L042102 (2024)], interesting dimensional crossovers [from two- to one-dimensional (2D to 1D) scaling] were found in the growth of Kardar-Parisi-Zhang (KPZ) interfaces on rectangular substrates, with lateral sizes $L_y > L_x$. Here, we extend this study to other universality classes for interface growth --- specifically, the Edwards-Wilkinson (EW), the Mullins-Herring (MH), and the Villain-Lai Das Sarma (VLDS) classes. From extensive simulations, we demonstrate that, in all systems with sufficiently large aspect ratio $\mathcal{R}=L_y/L_x$, the roughness $W$ scales with time $t$ in the growth regime as $W \sim t^{\beta_{\text{2D}}}$ for $t \ll t_c$ and $W \sim t^{\beta_{\text{1D}}}$ for $t \gg t_c$, where $t_c \sim L_x^{z_{2\text{D}}}$ in most cases. For the VLDS class, this crossover is also observed in the height distribution (HD), which approaches its characteristic probability density function for the 2D case at short times ($t \ll t_c$) and then crosses over to the asymptotic 1D HD. Dimensional crossovers are also found in the steady state regime, both in the roughness scaling as well as in the VLDS HD, which interpolate between the 2D and 1D ones as $\mathcal{R}$ increases. The particular case $L_x = L_y^{\delta}$, with $0 < \delta < 1$, is also discussed in detail and reveals interesting features of the investigated systems. For instance, there exist a `special' exponent $\delta^* = z_{1\text{D}}/z_{2\text{D}}$ such that the temporal crossover cannot be observed for $\delta > \delta^*$. Moreover, this leads the saturation roughness to display a nonuniversal scaling: $W_s \sim L_y^{\Lambda}$, with $\Lambda = (1-\delta) \alpha_{1\text{D}} + \delta \alpha_{2\text{D}}$.
\end{abstract}
\maketitle

\section{Introduction}
\label{secIntro}

Crossover phenomena are ubiquitous in nature. One of the most prominent examples is the smooth variation between two equilibrium phases of a given $d$-dimensional ($d$D) system, when its thermodynamic parameters are changed, without undergoing a phase transition \cite{CrossPhen}. Temporal crossovers are also widely observed between transient and long-time regimes. Here, we are interested in dimensional crossovers, where the system displays the behavior of different dimensionalities depending on its parameters or geometry. We remark that such dimensional crossovers are a subject of much interest in the literature, being theoretically/numerically investigated in various spins models --- as a consequence of either anisotropic couplings \cite{Graim,*Hatta,*Yamagata,*Lee,*Gonzalez} or of the system (nonequilateral) geometry \cite{Binder74,*Janke,*Laosiritaworn,*Pavel,*Pavel2,*Djordje,*Yusuf} ---, in heat conduction on rectangular domains \cite{PG1,*PG2,*Kiminori}, Bose-Einstein condensates \cite{Stein}, quantum phase transitions \cite{Qinpei}, confined solids \cite{Sun} and so on. Moreover, experimental examples include ultra-thin magnetic films \cite{Li,*Willis}, polymer coatings \cite{Sung}, layered superconductors \cite{Ruggiero,*Uji}, thermal \cite{Ghosh,*Navid,*Dong} and quantum transport \cite{Gehring,*Ali}, bosonic gases of trapped ultracold atoms \cite{Vogler,*Giulio,*Shah,*Yanliang} or photons \cite{Kirankumar}, etc.

In a recent work \cite{Ismael24}, dimensional crossovers were also observed in Kardar-Parisi-Zhang (KPZ) \cite{KPZ} growth on rectangular substrates (with lateral sizes $L_x \times L_y$). For instance, during the growth regime the surface roughness, $W$, was found to increase in time, $t$, as $W \sim t^{\beta_{2\text{D}}}$ and $W \sim t^{\beta_{1\text{D}}}$ at short and long times, respectively, provided that $L_y \gg L_x > 1$. Here, $\beta_{d\text{D}}$ is the KPZ growth exponent for substrate dimension $d$ \cite{Ismael24}. Hence, the rectangular substrates lead the system to undergo a KPZ$_{2\text{D}}$--KPZ$_{1\text{D}}$ crossover, from an initial 2D dynamics to an asymptotic 1D one. This behavior (expected for $L_y \gg L_x > 1$) can be summarized in the scaling relation \cite{Ismael24}:
\begin{equation}
 W(L_x,t) \simeq A^{\frac{1}{2}} L_x^{\alpha_{2\text{D}}} \mathcal{F}\left(\frac{bt}{L_x^{z_{2\text{D}}}}\right),
 \label{eqCrossScaling}
\end{equation}
where $\alpha_{2\text{D}}$ and $z_{2\text{D}}$ are the universal roughness and dynamic 2D KPZ exponents, while $A$ and $b$ are nonuniversal (system-dependent) parameters. Moreover, the scaling function follows $\mathcal{F}(x) \sim x^{\beta_{2\text{D}}}$ for $x \ll 1$ and $\mathcal{F}(x) \sim x^{\beta_{1\text{D}}}$ for $x \gg 1$. As explained in \cite{Ismael24}, initially, the system exhibits the expected two-dimensional behavior because the lateral correlation length $\xi \sim t^{1/z_{2\text{D}}}$ increases in both substrate directions. However, when it attains the smallest size $L_x$ (i.e., when $\xi \sim L_x$), at the crossover time $t_c \sim L_x^{z_{2\text{D}}}$, the fluctuations stop increasing in the $x$-direction and the system passes to behave as if it was one-dimensional. This crossover appears also in the height distributions (HDs) for the growth regime, where two scenarios were found depending on the system geometry: crossover from flat 2D HD to flat 1D HD; or from cylindrical 2D HD to circular 1D HD \cite{Ismael24}.

A dimensional crossover was observed also in the steady state regime (attained when $\xi \sim L_y > L_x$), where the saturation roughness was found to scale as \cite{Ismael24}:
\begin{equation}
 W_s^2(L_x,L_y) \simeq A L_x^{\alpha_{2\text{D}}} L_y^{\alpha_{2\text{D}}} \mathcal{G}\left( \frac{L_y}{L_x} \right),
 \label{eqScalingSS}
\end{equation}
with $\mathcal{G}(\mathcal{R}) \approx const.$ for $\mathcal{R} \approx 1$ and $\mathcal{G}(\mathcal{R}) \sim \mathcal{R}^{2\alpha_{1\text{D}}-\alpha_{2\text{D}}}$ for $\mathcal{R} \gg 1$. We notice that this scaling can alternatively be written in the simplified form:
\begin{equation}
 W_s^2 \simeq A L_x^{2 \alpha_{2\text{D}}} \mathcal{H}\left( \frac{L_y}{L_x} \right),
 \label{eqScalingSS2}
\end{equation}
with $\mathcal{H}(\mathcal{R}) \sim \mathcal{R}^{\alpha_{2\text{D}}} \sim const.$ for $\mathcal{R} \approx 1$ and $\mathcal{H}(\mathcal{R}) \sim \mathcal{R}^{2\alpha_{1\text{D}}}$ for $\mathcal{R} \gg 1$. The steady state regime HDs also present a dependency on the substrate aspect ratio $\mathcal{R}=L_y/L_x$, interpolating between the 2D HD, for $\mathcal{R} \rightarrow 1$, and the 1D HD as $\mathcal{R} \rightarrow \infty$ \cite{Ismael24}.

We remark that several experimental techniques for selective area growth --- where deposition occurs inside predefined regions of the substrate --- have been recently developed to produce rectangular nanosheets/nanowalls, horizontal nanowires, etc. \cite{Chi,*Schmid,*Murillo,*Winnerl}. So, given the appeal of these nanostructures for a wide variety of applications \cite{Yuan,*Wang}, it is very important to deeply understand how the rectangular substrate affects growth processes in general, i.e., beyond the KPZ case. In order to do this, we present here a detailed analysis of models belonging to the linear classes by Edwards-Wilkinson (EW) \cite{EW} and Mullins-Herring (MH) \cite{Mullins,*Herring}, as well as the non-linear class by Villain-Lai-Das Sarma (VLDS) \cite{Villain,*LDS} deposited on 2D substrates of lateral sizes $L_y > L_x$. In a brief, we find that the same crossover scalings for the KPZ case (Eqs. \ref{eqCrossScaling} and \ref{eqScalingSS}) also apply for these other classes. The only exception is the EW class, where $W$ displays logarithmic behaviors in 2D; and the appropriate scaling relations for this case are devised here. The crossover in the HDs is also discussed for these classes, in both the growth and saturation regimes. We investigate also the interesting situation where $L_x = L_y^{\delta}$ (including KPZ systems), demonstrating that the temporal crossover cannot happen for $\delta > \delta^* = z_{1\text{D}}/z_{2\text{D}}$. Substantially, the $L_x = L_y^{\delta}$ condition uncovers a non-universal scaling $W_s \sim L_y^{\Lambda}$, with $\Lambda = (1-\delta) \alpha_1 + \delta \alpha_2$, for the saturation roughness.

The remainder of the paper is organized as follows. The investigated models and quantities of interest are defined in Sec. \ref{secModels}. Results for the growth and steady state regimes are presented in Secs. \ref{secGR} and \ref{secSSR}, respectively. The particular case with $L_x = L_y^{\delta}$ is analyzed in Sec. \ref{secPG}. Section \ref{secConc} presents our final discussions and conclusion.

\section{Models and quantities of interest}
\label{secModels}

\subsection{Models}

We performed extensive Monte Carlo simulations of several discrete growth models on rectangular square lattice substrates of lateral sizes $L_x \times L_y$, with periodic boundary conditions in both $x$ and $y$ directions.

In the KPZ class, we studied the restricted solid-on-solid (RSOS) \cite{KK} and the Etching model \cite{Mello01}. The RSOS model with deposition and evaporation (RSOSev) \cite{tiago06} and the Family model \cite{Family} are representatives of the EW class. We also investigate large curvature (LC) models \cite{LCM,KrugLCM} and the conserved RSOS (CRSOS) model \cite{CRSOS}, which belong to the MH class and VLDS class, respectively.

In all cases, particles are sequentially deposited at random positions of the substrate and, once a site $(i,j)$ --- whose four nearest neighbors (NNs) will be denoted here as $\partial_{ij}$ --- is chosen, the aggregation rules there are as follows:

\begin{itemize}

 \item \textit{RSOS} \cite{KK}: $h_{ij} \rightarrow h_{ij} +1$ if $|\Delta h| \leq m$, after deposition, $\forall$ NNs $\partial_{ij}$; otherwise, the particle is rejected. $m$ is a positive integer parameter and $\Delta h \equiv h_{ij} - h_{\partial_{ij}}$.

\item \textit{Etching} \cite{Mello01}: $h_{\partial_{ij}} \rightarrow \max[h_{ij}, h_{\partial_{ij}}]$  $\forall$ NNs $\partial_{ij}$ and, then, $h_{ij} \rightarrow h_{ij} +1$.

\item \textit{RSOSev} \cite{tiago06}: With equal chance, we first choose the next event (a deposition $h_{ij} \rightarrow h_{ij} +1$ or an evaporation $h_{ij} \rightarrow h_{ij} -1$) to be performed. Such an event only occurs if the RSOS constraint ($|\Delta h| \leq 1$, $\forall$ NNs $\partial_{ij}$) is satisfied after the deposition or evaporation; otherwise, the event attempt is rejected.

\item \textit{Family} \cite{Family}: If $h_{ij}\leq h_{\partial_{ij}}$ $\forall$ NNs $\partial_{ij}$, then $h_{ij} \rightarrow h_{ij} +1$; otherwise, the particle moves to the NN site with minimal height, with a random draw resolving possible ties.

\item \textit{LC1} \cite{LCM}: The local curvature $C_{k} \equiv \nabla^4 h$ is calculated around position $k$ and, if $C_{ij}\leq C_{\partial_{ij}}$ $\forall$ NNs $\partial_{ij}$, then $h_{ij} \rightarrow h_{ij} +1$; otherwise, the particle moves to the NN site with the largest $C_{\partial_{ij}}$, with a random draw resolving possible ties.

\item \textit{LC2} \cite{KrugLCM}: The rule is identical to the LC1 model, but instead of selecting a lattice site $(ij)$, an interstitial position $(i+1/2,j+1/2)$ is randomly chosen and deposition occurs at its adjacent site with the largest curvature.

 \item \textit{CRSOS} \cite{CRSOS}: The freshly deposited particle can diffuse at the surface until finding a site [let us say, $(i',j')$] where the RSOS constraint above is satisfied; then, $h_{i'j'} \rightarrow h_{i'j'} +1$.

\end{itemize}

For the RSOS model, only the original version, with $m=1$, will be analyzed here. On the other hand, the CRSOS model is studied for $m=1$ and $m=4$; and we will refer to these models as CRSOS1 and CRSOS4. In all cases, the growth is performed on initially flat substrates, i.e., $h_{ij}(t=0)=0$ for all sites $i,j$. Moreover, the number $N$ of samples grown in each case was such that $N L_x L_y \gtrsim 10^{8}$.

\subsection{Quantities of interest}

Once we have a height field $h_{ij}(t)$ [generated by one of the models above], its fluctuations can be quantified through the central moments:
\begin{equation}
 M_n(t) = \langle \overline{[h_{i,j}(t) - \bar{h}(t)]^n} \rangle,
\end{equation}
where $\bar{\cdot}$ represents spatial average over all $L_x\times L_y$ substrate sites and $\langle \cdot \rangle$ denotes the configurational average over the $N$ different samples.

The main observable used to investigate a given growth process is the global surface roughness (or width):
\begin{equation}
W = \sqrt{M_2},
\end{equation}
which is the standard deviation of the height distribution (HD). To obtain a more complete characterization of such HD, we can also study adimensional moment ratios, such as the skewness ($S$):
\begin{equation}
S = \frac{M_3}{M_2^{3/2}},
\end{equation}
and the excess kurtosis ($K$):
\begin{equation}
K = \frac{M_4}{M_2^2}-3,
\end{equation}
which respectively quantify the asymmetry and the weight of the tails of a given probability distribution.

Following Ref. \cite{Ismael24}, it is interesting to also calculate the ``line moments'':
\begin{equation}
 M_n^{(\ell)}(t) = \langle \overline{[h_{i,j}(t) - \bar{h}(t)]^n} \rangle_{\ell},
\end{equation}
where the spatial average is performed only in the $\ell (= x$ or $y)$ direction. Therefore, $M_n^{(x)}(t)$ and $M_n^{(y)}(t)$ measure the  fluctuations in the respective substrate direction. Our main interest here will be in the ``line roughness'' $W_{x} = \sqrt{M_2^{(x)}}$ and $W_{y} = \sqrt{M_2^{(y)}}$.

\section{Results for the growth regime}
\label{secGR}

In this section, we will focus on the growth regime. Therefore, very long substrate sizes ($L_y=32768$ for the EW models and $L_y=8192$ in the other cases) will be considered in the $y$-direction, in order to avoid the saturation regime within the times analyzed.

\subsection{Global roughness}

\begin{figure}[t]
 \includegraphics[width=4.2cm]{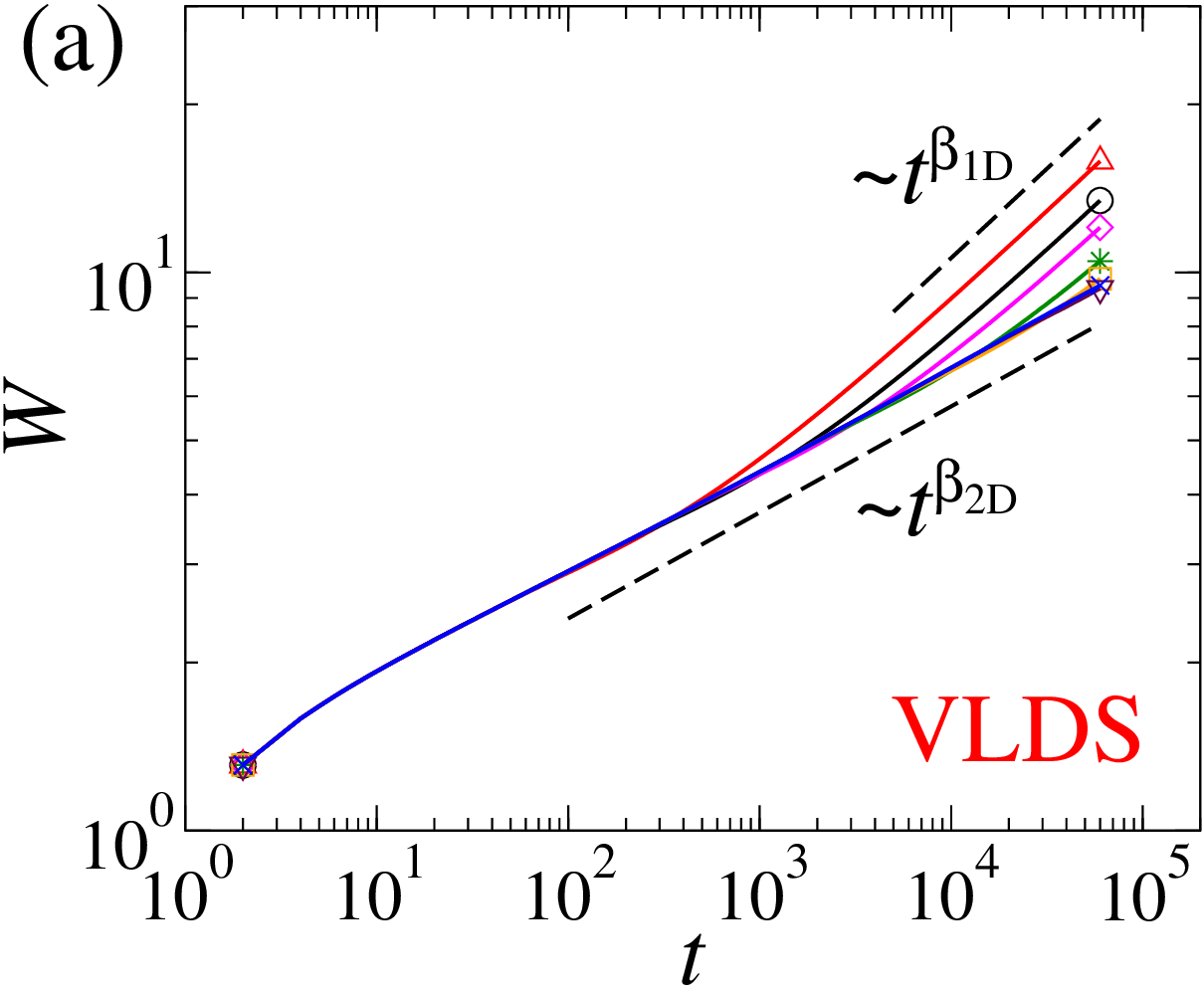}
 \includegraphics[width=4.2cm]{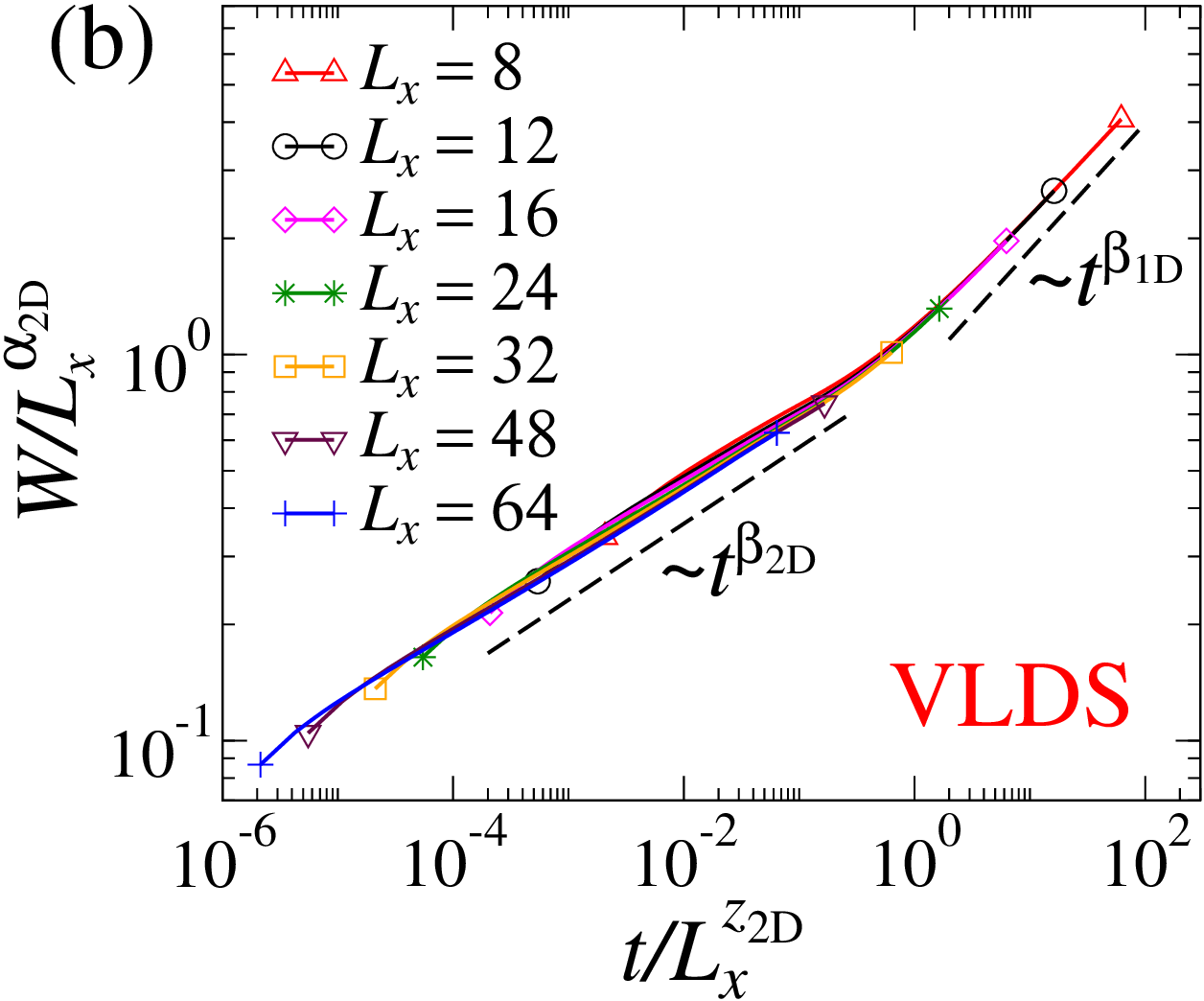}
 \includegraphics[width=4.2cm]{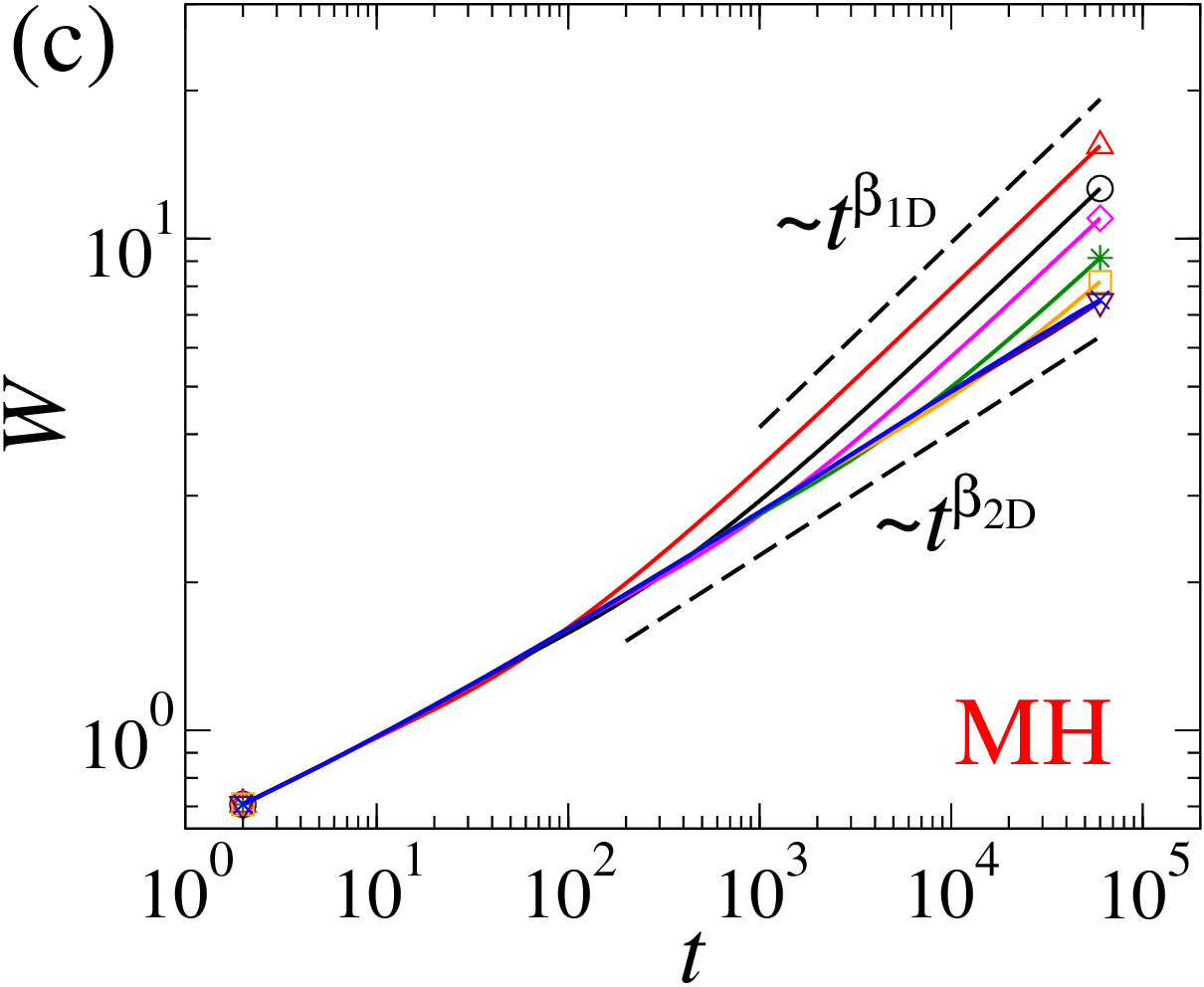}
 \includegraphics[width=4.2cm]{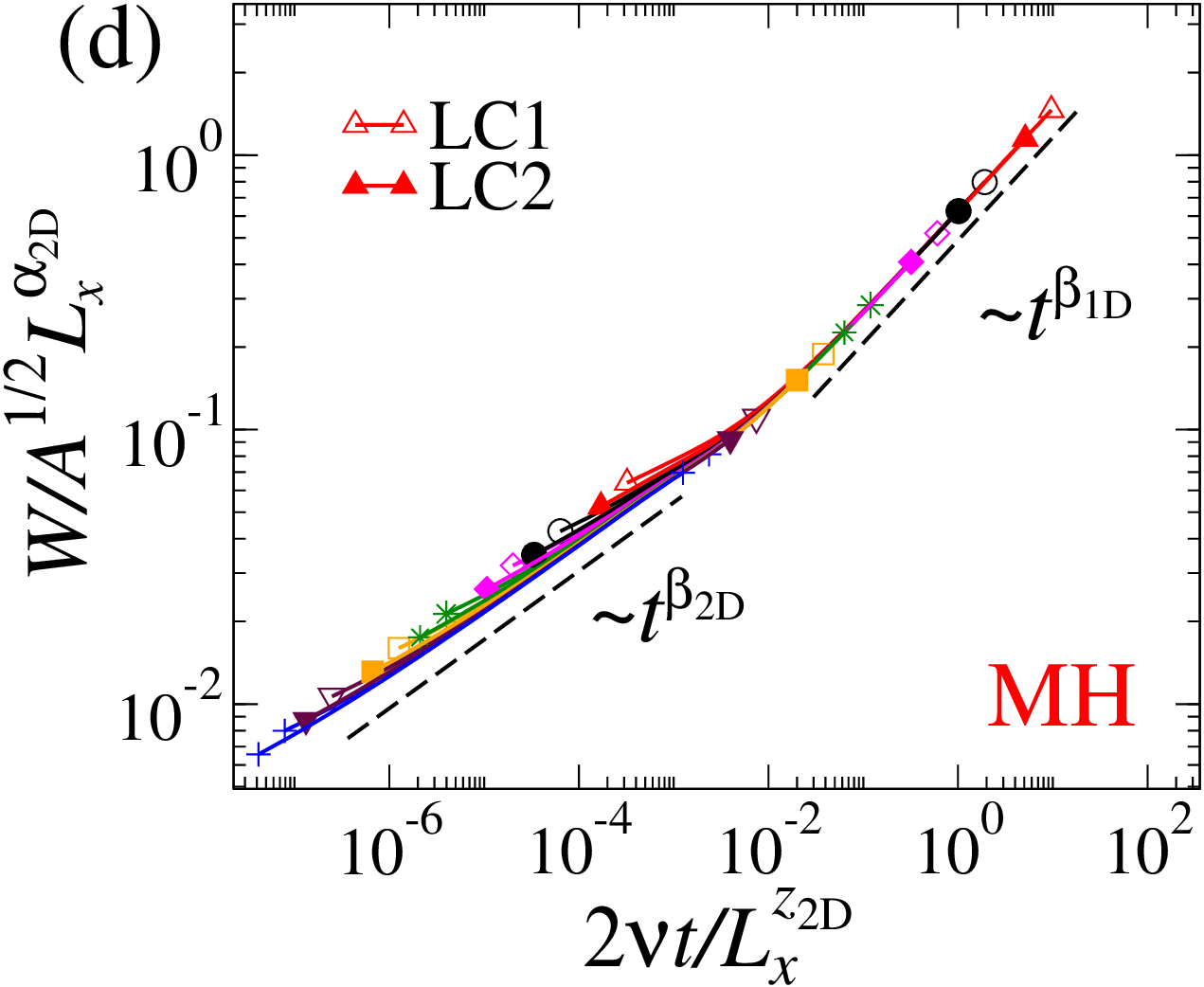}
 \includegraphics[width=4.2cm]{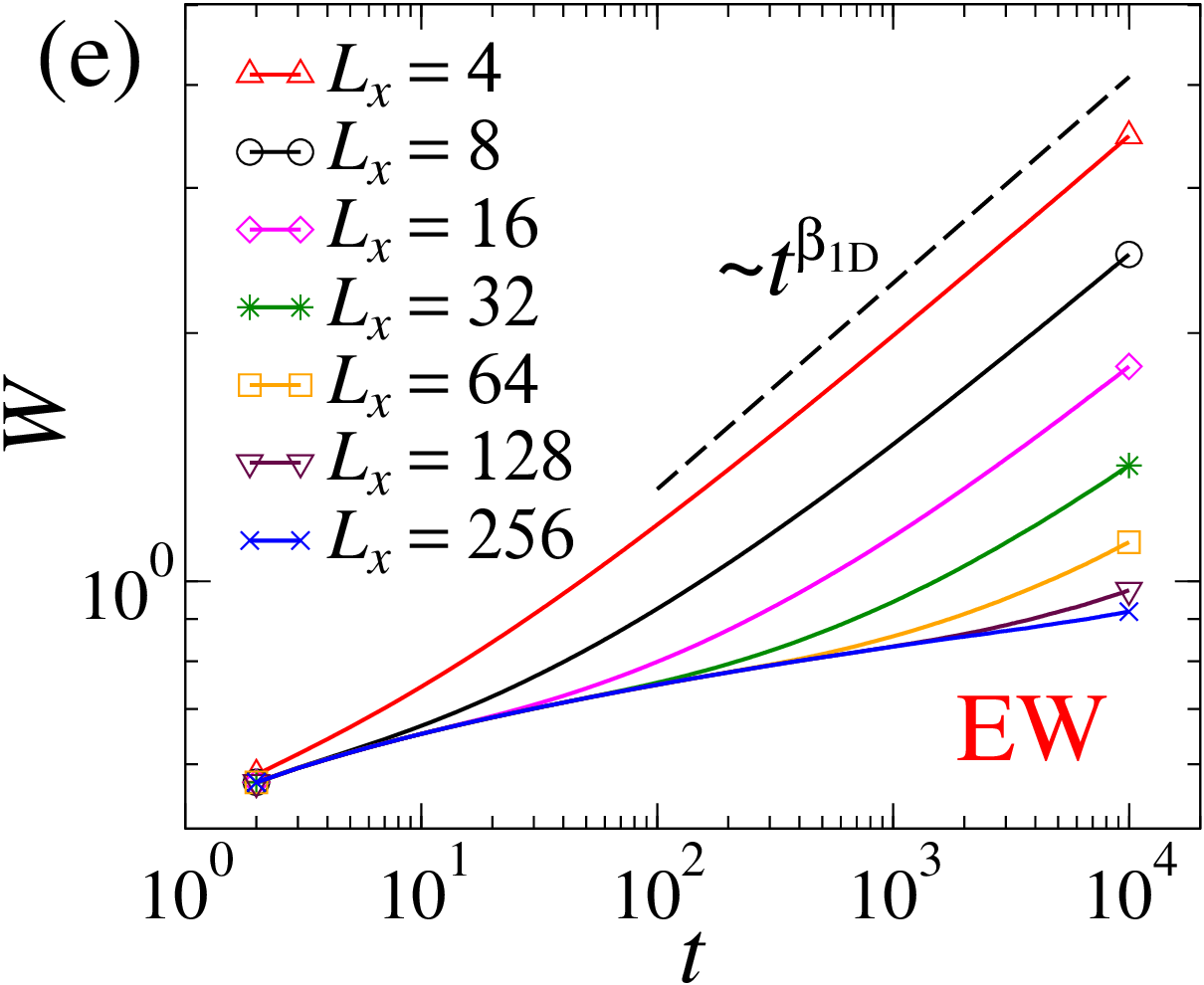}
 \includegraphics[width=4.2cm]{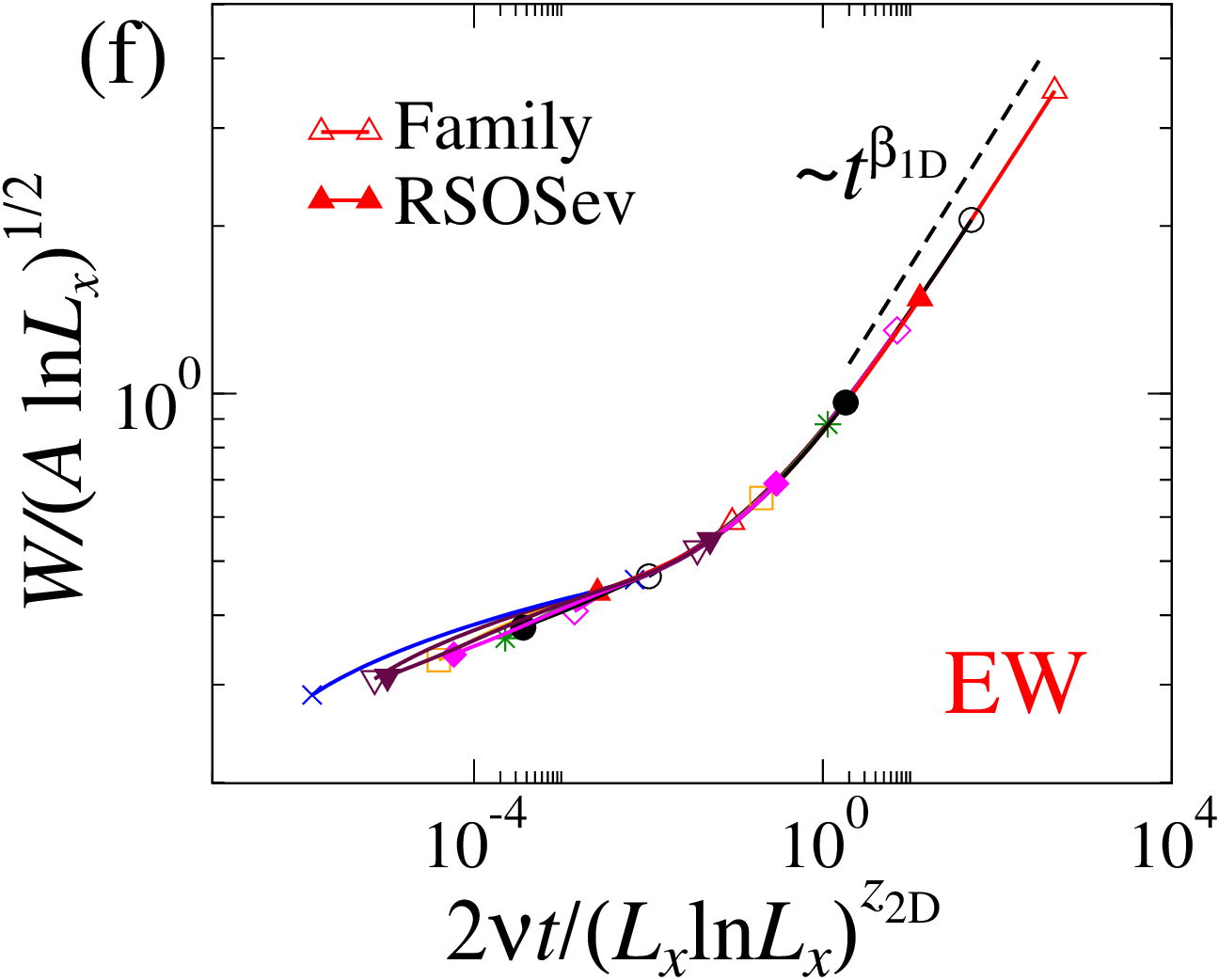}
 \caption{Global roughness $W$ versus time $t$ (left) and corresponding rescaled curves (right panels) for the CRSOS4 [(a) and (b)], LC2 [(c) and (d)], and Family model [(e) and (f)]. Results for several lateral sizes $L_x$ are presented, as indicated by the legends, whereas $L_y=32768$ for the EW models and $L_y=8192$ for the other ones. The sizes for the LC models are the same as in (b). Moreover, open and closed symbols represent data for different models in (d) and (f). The dashed lines have the indicated slopes, with the values of $\beta_{d\text{D}}$ (and also of $z_{d\text{D}}$) for each class being given in the text.}
\label{fig1}
\end{figure}

Figure \ref{fig1} presents the temporal variation of the global roughness, $W$, for models in the VLDS, MH, and EW classes. Similarly to the behavior observed in \cite{Ismael24} for KPZ systems, in all cases $W$ starts increasing according to the 2D scaling at short times, but then, at a time $t_c$, it crosses over to an asymptotic 1D regime.

We recall that the exact exponents for the EW and MH classes are $\alpha_{d\text{D}}=\dfrac{2\kappa-d}{2}$, $\beta_{d\text{D}}=\dfrac{2\kappa-d}{4\kappa}$ and $z_{d\text{D}}=2\kappa$, with $\kappa=1$ and 2 in the EW and MH case, respectively \cite{barabasi,KrugAdv}. The two-loop exponents for the VLDS class are $\alpha_{d\text{D}} = \dfrac{4-d}{3}-\epsilon$, $\beta_{d\text{D}} = \dfrac{4 -d - 3\epsilon}{8+d-6\epsilon}$ and $z_{d\text{D}} = \dfrac{8+d}{3}-2\epsilon$, with $\epsilon =  0.01361(2 - d/2)^2$ \cite{Janssen}. Therefore, for the MH and VLDS classes (as well as for KPZ systems \cite{Ismael24}), the dimensional crossover happens between two power-law regimes: $W \sim t^{\beta_{2\text{D}}}$ for $t \ll t_c$ and $W \sim t^{\beta_{1\text{D}}}$ for $t \gg t_c$, as seen in Figs. \ref{fig1}(a)-\ref{fig1}(d).

In contrast with these classes, in the 2D EW case the roughness does not increase algebraically, presenting instead a logarithmic behavior in time, $W^2 \sim \ln t$, in the growth regime (since $\beta_{2\text{D}}=0$). In one-dimension, on the other hand, we have the scaling $W \sim t^{1/4}$ for EW systems \cite{barabasi,KrugAdv}. Therefore, the temporal 2D-to-1D crossover in this class is expected to happen from an initial slow (logarithmic) variation of $W$ to an asymptotic power-law increase, as indeed observed in Figs. \ref{fig1}(e) and \ref{fig1}(f).

In all cases, the crossover time $t_c$ is an increasing function of $L_x$. For the VLDS and MH classes, we find that $t_c \sim L_x^{z_{2\text{D}}}$, similarly to the behavior found in \cite{Ismael24} for KPZ models. In fact, by rescaling the curves of $W\times t$ for different $L_x$ according to the scaling relation $\ref{eqCrossScaling}$, a good data collapse is observed in Figs. \ref{fig1}(b) and \ref{fig1}(d) respectively for the VLDS and MH systems. As shown in Ref. \cite{Ismael24}, by taking into account the non-universal parameters $A$ and $b$ in Eq. \ref{eqCrossScaling}, even data from different models can be collapsed by using the 2D system values of $A$ and $b$. However, such parameters are not available for the 2D VLDS models and, as discussed in Ref. \cite{Ismael16a}, it is not clear how they can be obtained. Hence, in this case, we are not able to appreciate the data collapse beyond a single model.

On the other hand, for the linear (MH and EW) classes, the non-universal parameters are given by $A=D/\nu$ and $b=2\nu$, where $D$ and $\nu$ are the coefficients of the MH and EW equations: $\partial_t h = \nu \nabla^{2\kappa} h + \sqrt{D} \eta$ \cite{KrugAdv}. As discussed in Ref. \cite{Ismael19}, we expect that $D=1/2$ for both LC models and the Family model. Moreover, $\nu_{\text{LC1}} = 0.33(1)$ and $\nu_{\text{LC1}}=0.176(5)$ were numerically estimated there in two-dimensions \footnote{There is a typo in Table III of Ref. \cite{Ismael19} and the $\nu$ values reported there for LC1 and LC2 models are exchanged.}, yielding $A_{\text{LC1}} = 1.52(5)$ and $A_{\text{LC2}}=2.84(8)$. Indeed, using such parameters, the curves for both LC models collapse very well, as shows Fig. \ref{fig1}(d).

Since $\alpha_{2\text{D}}=0$ for 2D EW systems, their saturation roughness increases logarithmically with the system size, $W_s^2 \sim \ln L$, whereas $W_s \sim L^{\alpha_{2\text{D}}}$ in the other classes. This indicates that the term $L_x^{\alpha_{2\text{D}}}$ in scaling relation \ref{eqCrossScaling} has to be replaced by $(\ln L_x)^{1/2}$ in the EW case. Moreover, Eq. \ref{eqCrossScaling} gives $W \sim t^{\beta_{1\text{D}}}/L_x^{z_{2\text{D}}\beta_{1\text{D}}-\alpha_{2\text{D}}}$ for $t \gg t_c$, suggesting that $W \sim t^{\beta_{1\text{D}}}/L_x^{z_{2\text{D}}\beta_{1\text{D}}}$ for EW systems. These considerations lead to the modified scaling

\begin{equation}
 W(L_x,t) \simeq (A \ln L_x)^{1/2} \mathcal{F}_{EW}\left[\frac{2 \nu t}{(L_x \ln L_x)^{2}}\right],
 \label{eqCrossScalingEW}
\end{equation}
valid for $L_y \gg L_x$, where $\mathcal{F}_{EW}(x) \sim x^{\beta_{1\text{D}}}$ for $x \gg 1$. Indeed, a very good collapse is observed in Fig. \ref{fig1}(f), by rescaling the data for different EW models following this relation. For the 2D Family model, the non-universal parameters $A \approx 0.71$ and $\nu \approx 0.70$ were reported in Ref. \cite{Ismael19}, while the values $A \approx 1.06$ and $\nu \approx 0.18$ were estimated here for the RSOSev model (see the Appendix). Interestingly, this scaling indicates that the crossover time in the EW case is given by $t_c \sim L_x^{z_{2\text{D}}} (\ln L_x)^2$. Namely, it gains a multiplicative logarithmic `correction', which is absent for the other classes.

\begin{figure}[t]
 \includegraphics[width=4.2cm]{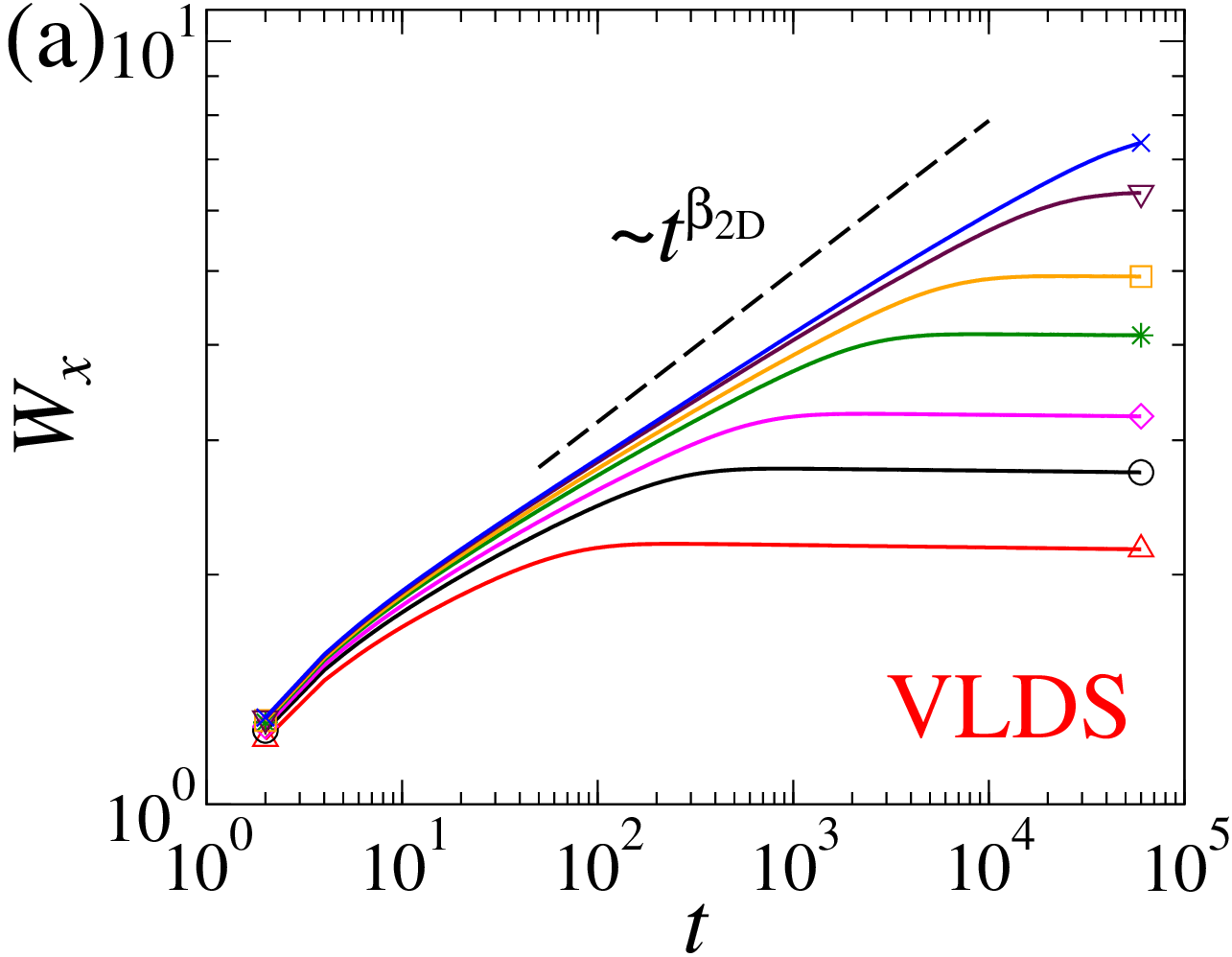}
 \includegraphics[width=4.2cm]{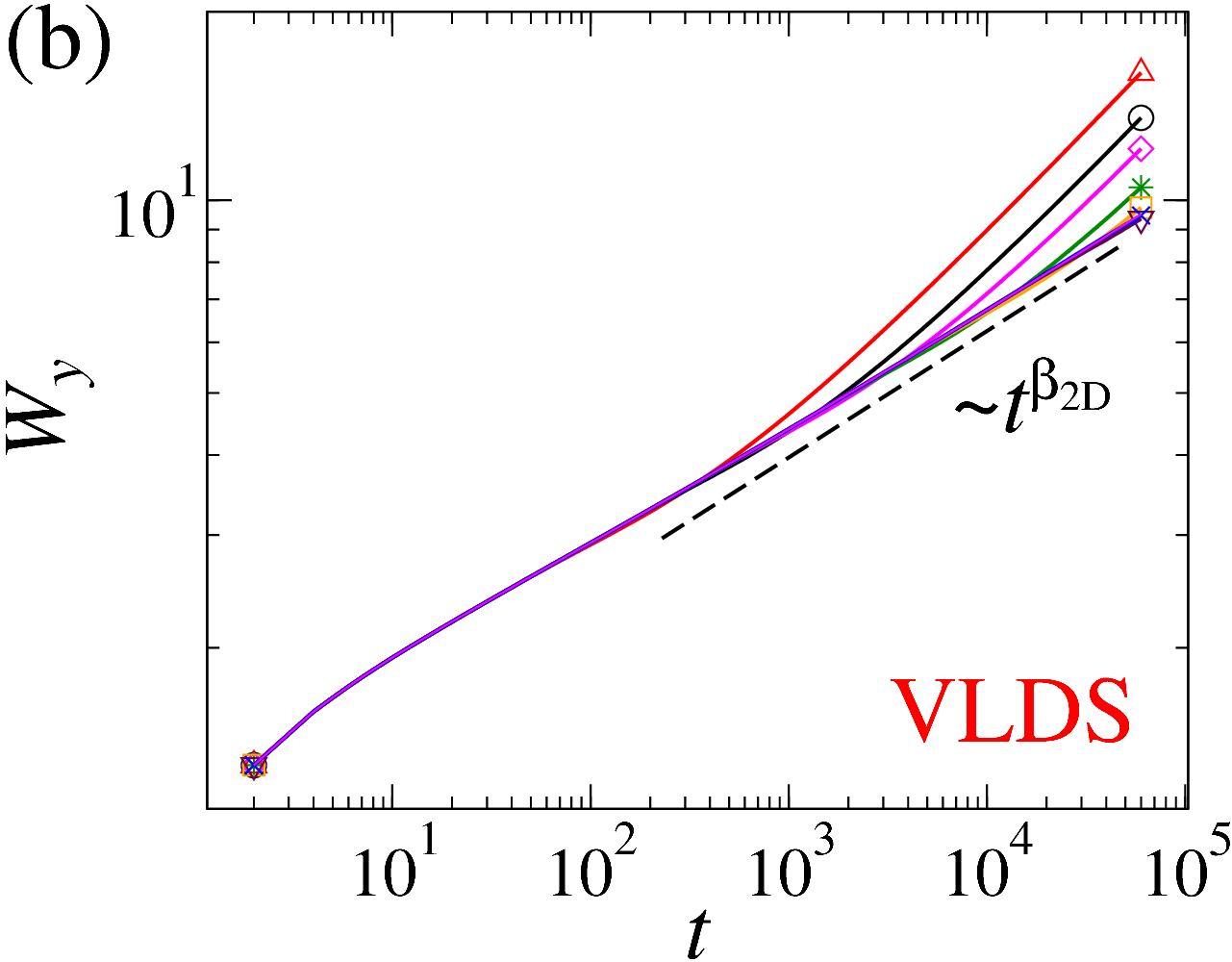}
 \includegraphics[width=4.2cm]{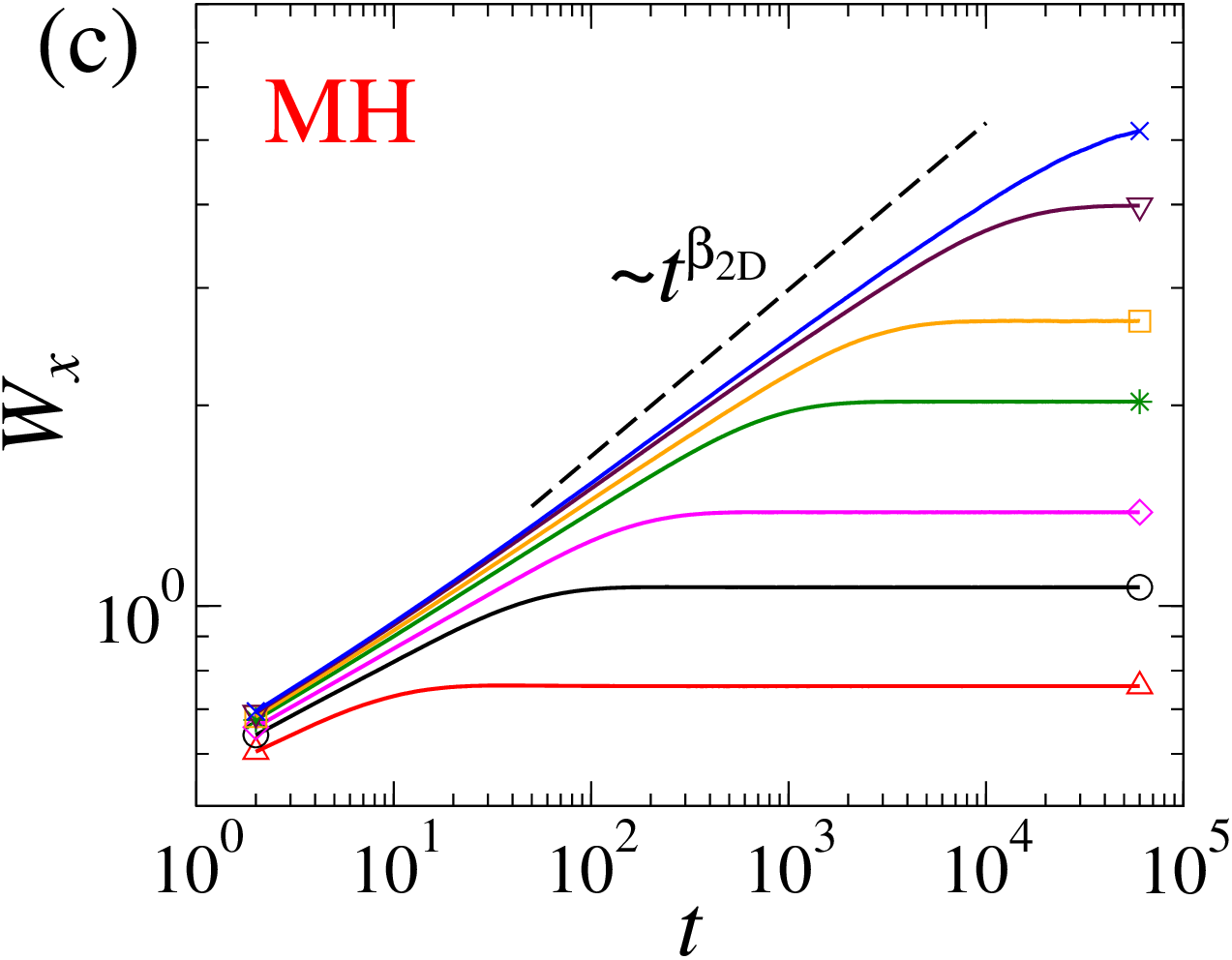}
 \includegraphics[width=4.2cm]{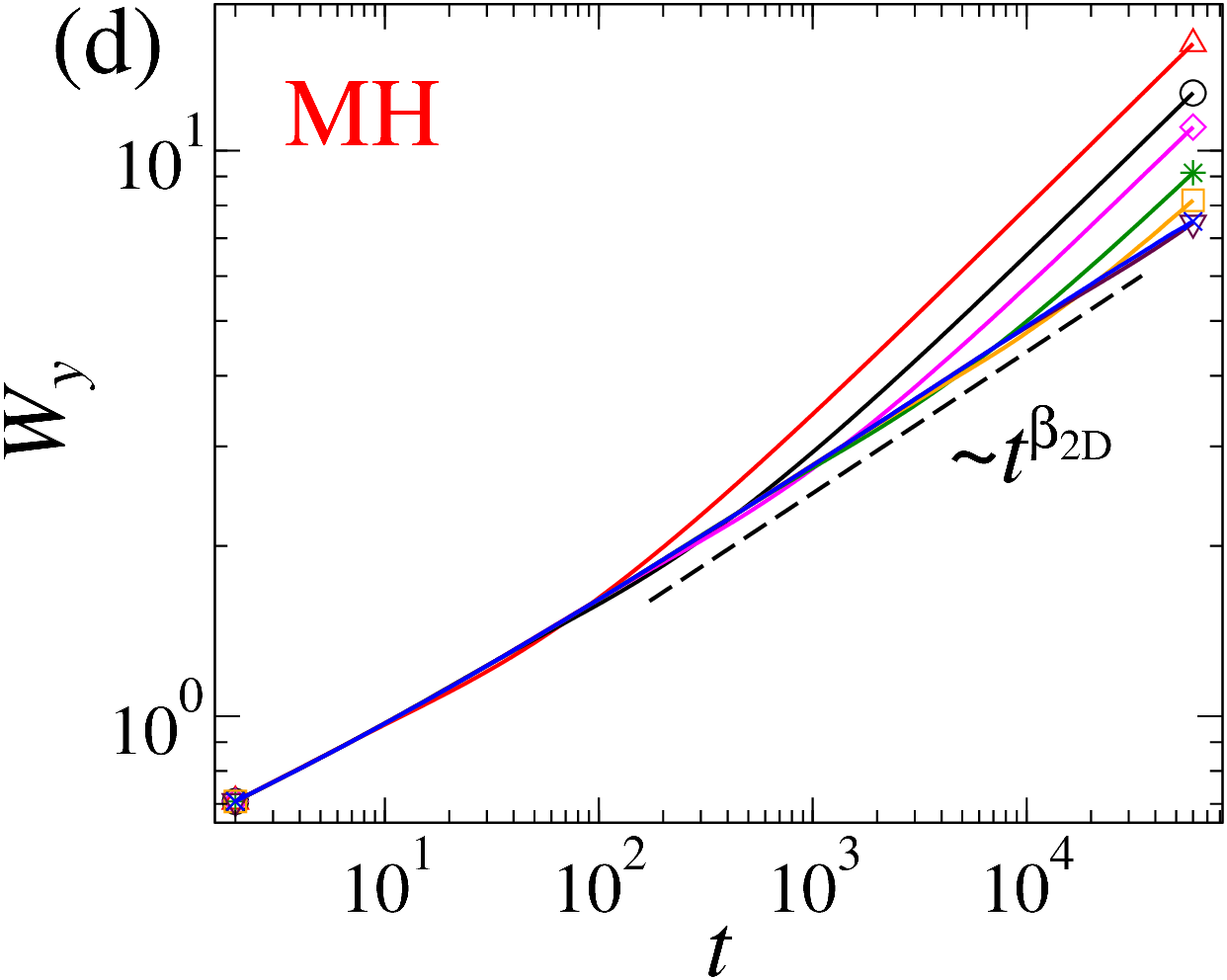}
 \includegraphics[width=4.2cm]{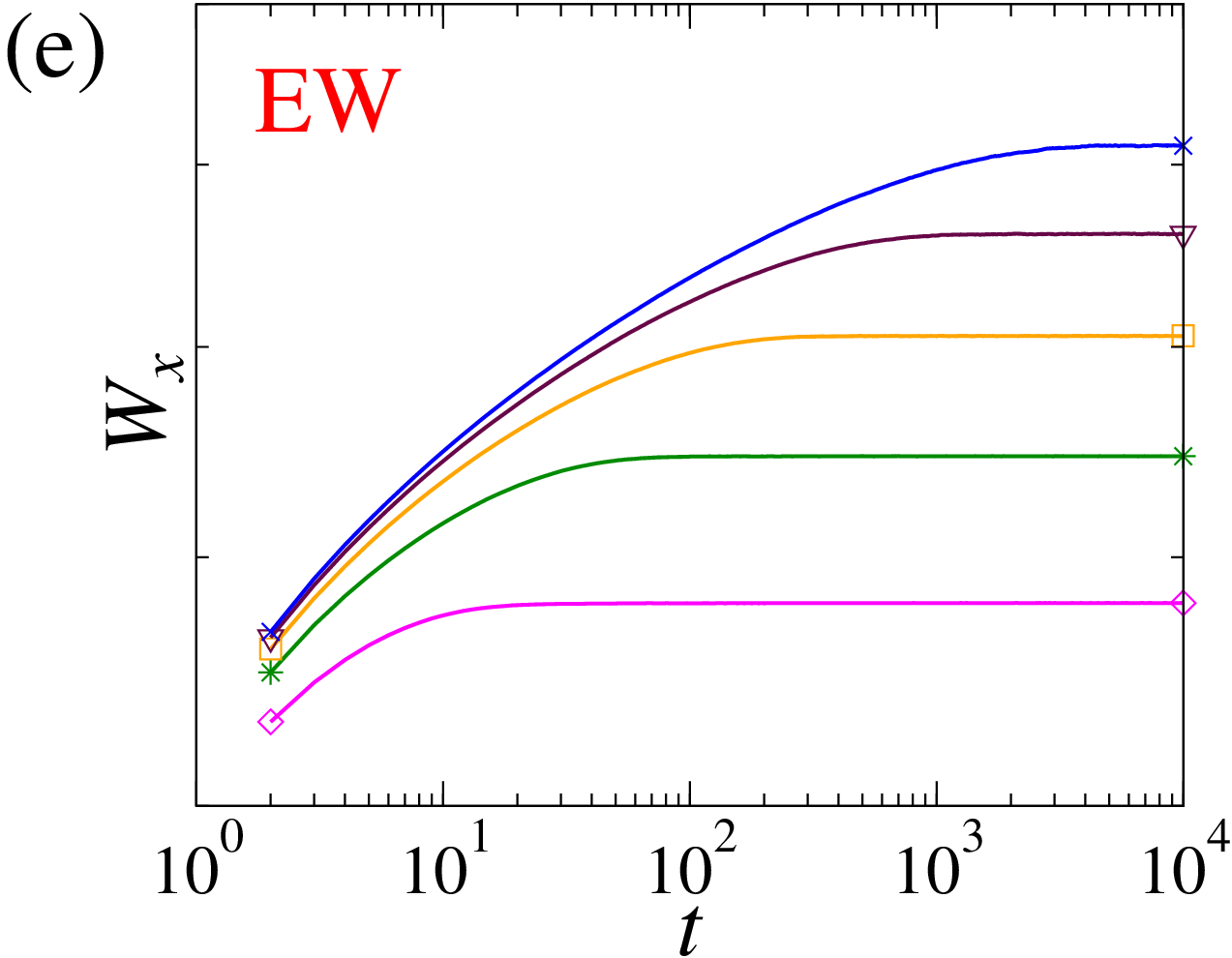}
 \includegraphics[width=4.2cm]{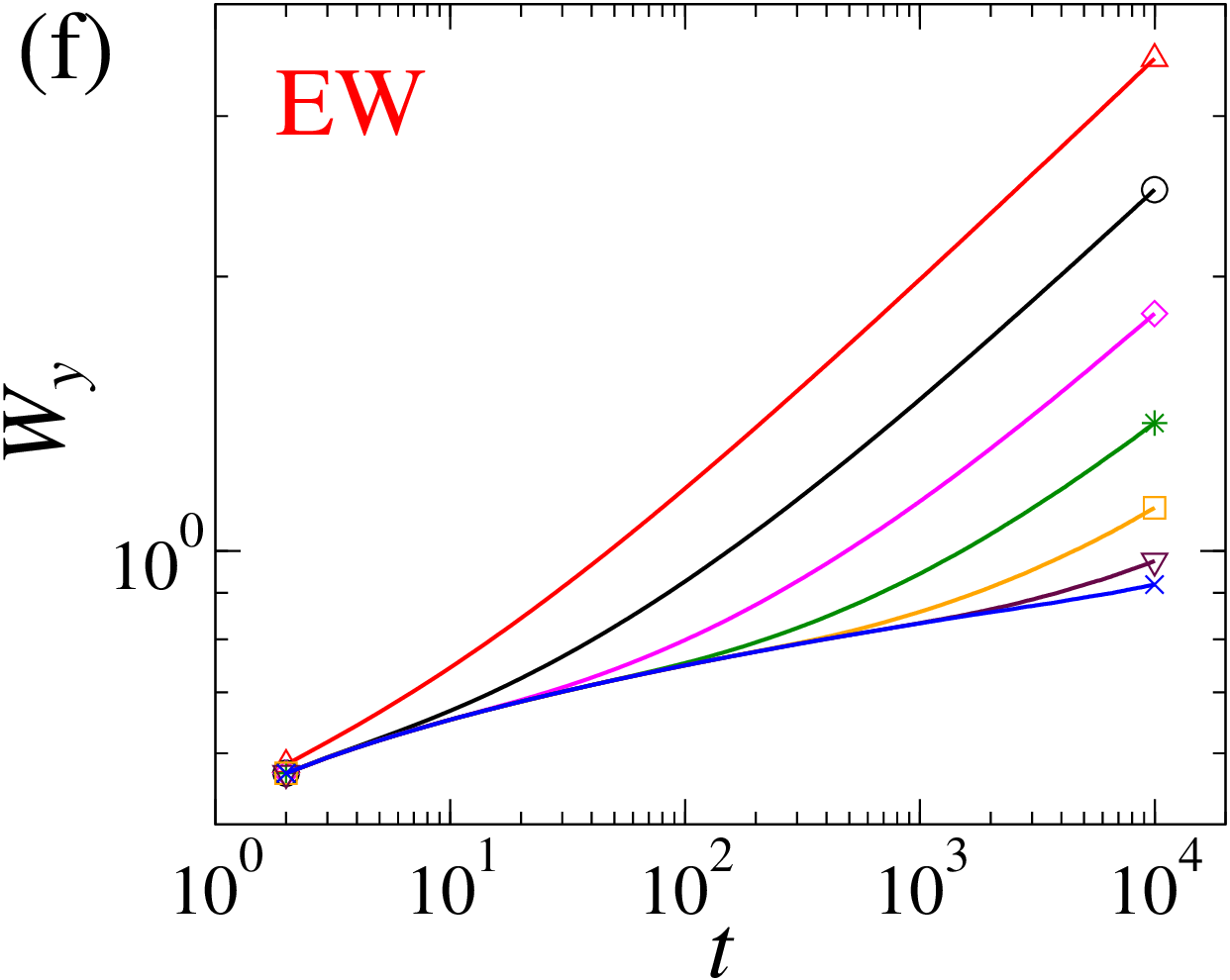}
 \caption{Line roughness $W_x$ (left) and $W_y$ (right panels) versus time $t$, for the CRSOS4 [(a) and (b)], LC2 [(c) and (d)], and Family model [(e) and (f)]. Data for several lateral sizes $L_x$ are presented, which are the same as those given by the legends in Fig. \ref{fig1}. The dashed lines have the indicated slopes, with the corresponding values of $\beta_{d\text{D}}$ of each class, as provided in the text.}
\label{fig2}
\end{figure}

\subsection{Line roughness}

To verify whether the origin of the dimensional crossover in the systems studied here is the same as that found for KPZ models in \cite{Ismael24}, it is interesting to analyze the directional roughness, $W_x$ and $W_y$, calculated along lines in the $x$ and $y$ directions, respectively. Figure \ref{fig2} shows the temporal variation of $W_x$ and $W_y$ for models in the three classes we are investigating. Analogously to the KPZ behavior, in all cases we find that both $W_x$ and $W_y$ increase at short times (i.e., for $t \ll t_c$) approximately as the global roughness of each class. Though stronger finite-size corrections are present in the scaling for $W_x$, because we are dealing with small $L_x (\ll L_y)$. Notably, the entire curves of $W_y \times t$ are very similar to those in Fig. \ref{fig1} for $W \times t$. On the other hand, for $t \gg t_c$, $W_x$ saturates, as a consequence of the correlation length $\xi$ becoming equal to $L_x$. Therefore, since the correlations stop increasing in the $x$-direction, but keep augmenting in the $y$-direction, the system passes to behave as if it was one-dimensional, explaining why $W_y$ (and then $W \approx W_y$) crosses over to an asymptotic 1D scaling.

\subsection{Height distributions}

As mentioned above, in the KPZ class the dimensional crossover is not limited to the roughness scaling, but it also manifests in the height distributions (HDs), which approach the probability density functions (PDFs) characteristic of two-dimensions at short times, but then change to the HDs of the 1D case. So, it is interesting to explore this also for the other classes.

We notice, however, that for the EW and MH classes the HDs are Gaussian in both one- and two-dimensions. Therefore, their skewness and kurtosis are null in both 2D and 1D regimes, so that no crossover is observed in such moment ratios. As demonstrated by some of us in \cite{Ismael19}, by appropriately taking into account the non-universal parameters in the scaling for the HDs, the Gaussians' variance for a given class assume universal values, which are different in each dimension. However, it seems not possible to use this fact here to observe a Gaussian-Gaussian crossover for the EW and MH classes, because the temporal scaling of $W$ is different at short and long times.

\begin{figure}[t]
 \includegraphics[width=4.2cm]{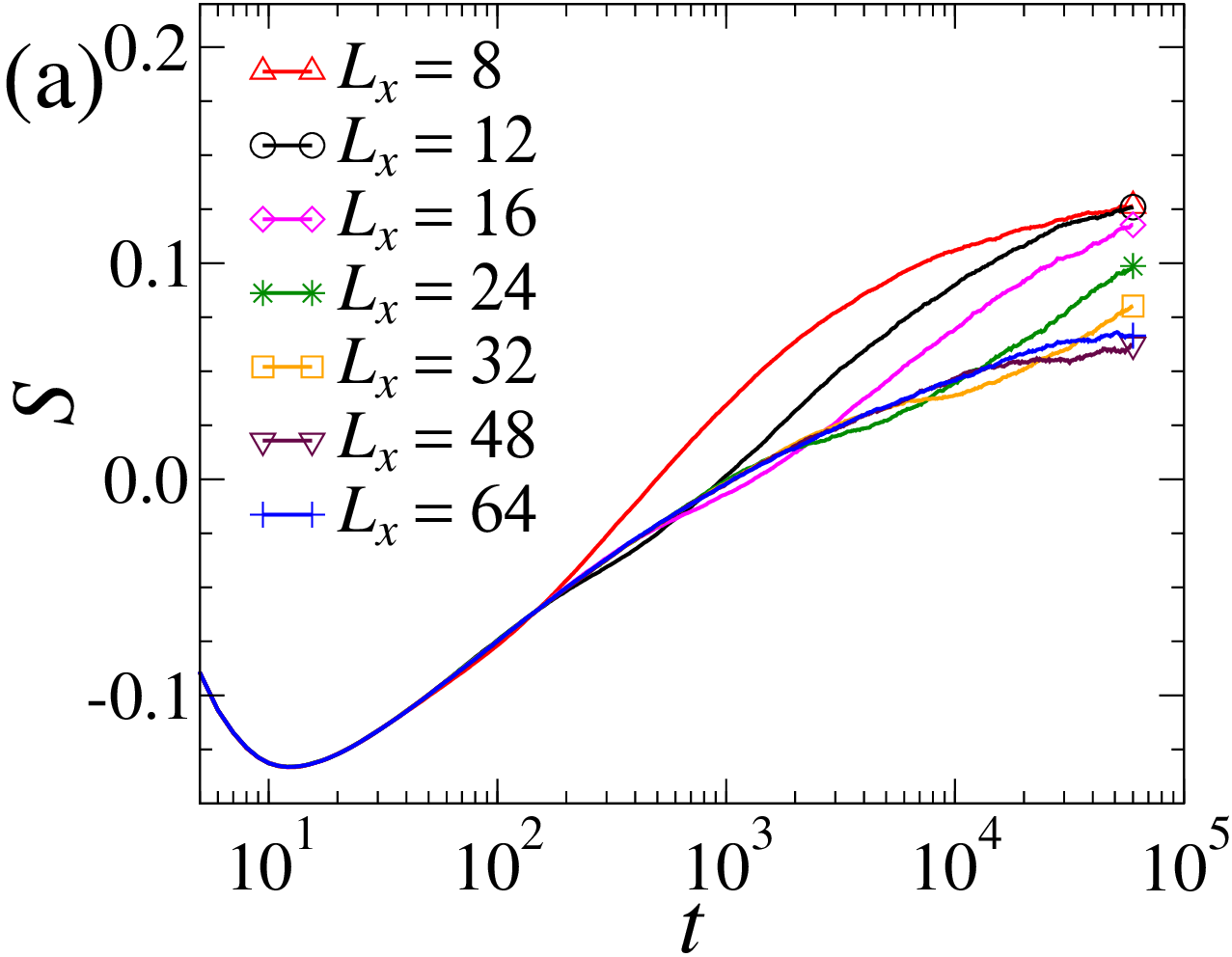}
 \includegraphics[width=4.2cm]{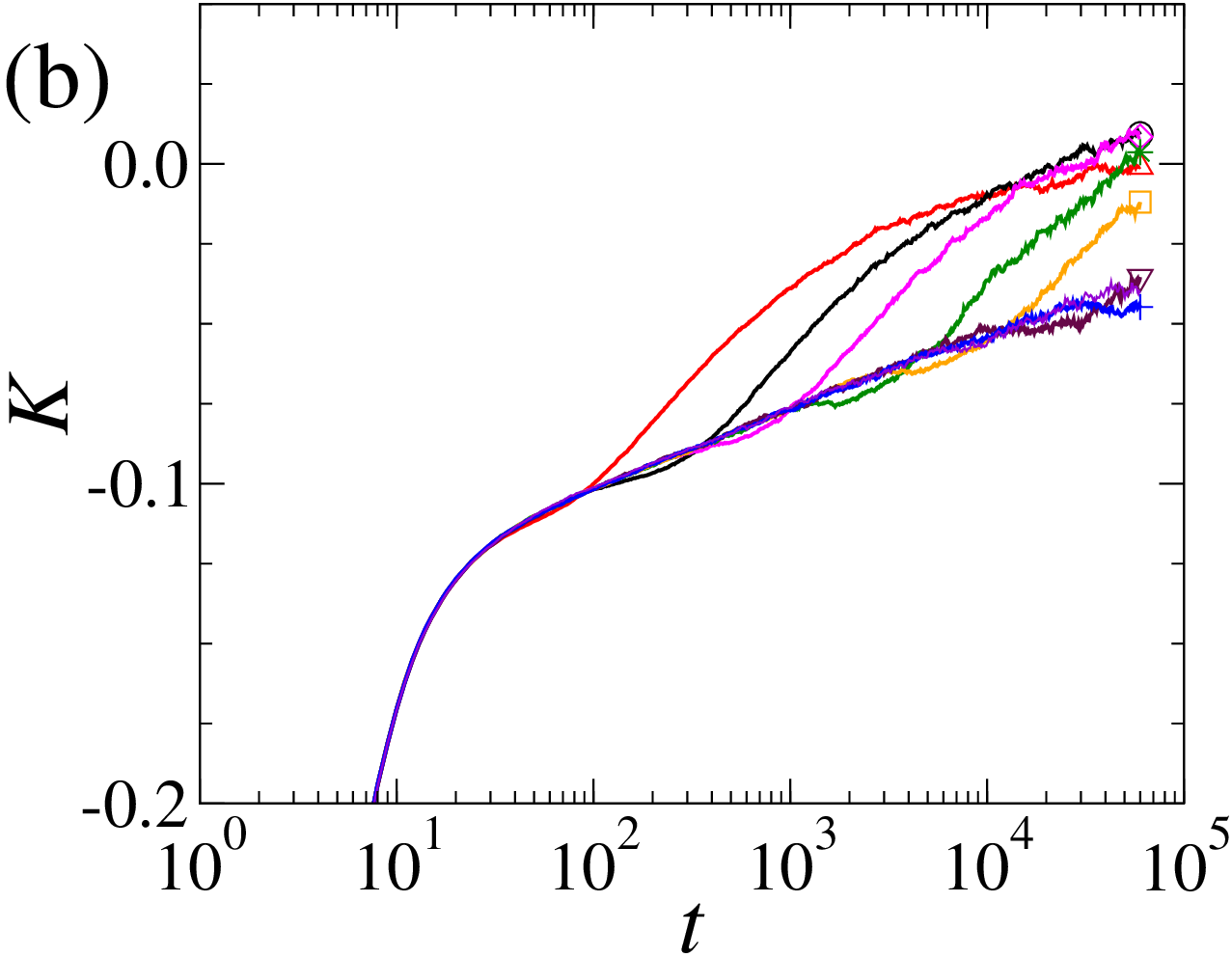}
 \caption{Temporal evolution of the (a) skewness $S$ and (b) kurtosis $K$ of the growth regime HD for the CRSOS4 model. The substrate sizes $L_x$ are indicated by the legend, whereas $L_y = 8192$.}
\label{fig3}
\end{figure}

A different scenario is expected in the VLDS case, since the HDs for this class are non-Gaussian \cite{Ismael16a}. Indeed, a clear crossover is observed in the temporal variation of the skewness and kurtosis in Fig. \ref{fig3}. It is worth mentioning that, although the crossover in this figure seems to suggest that the ratios for the 1D and 2D HDs are converging to very different places, essentially the same asymptotic values ($S \approx 0.13$ and $K \approx 0$) were found for them in Ref. \cite{Ismael16a}. Anyway, there exist a crossover in the VLDS HDs, which affects at least their convergence.

\section{Results for the steady state regime}
\label{secSSR}

We now investigate the saturation regime, where the lateral correlation length has reached the largest substrate dimension (i.e., $\xi \sim L_y>L_x$).

\subsection{Roughness scaling}

\begin{figure}[t]
 \includegraphics[width=4.2cm]{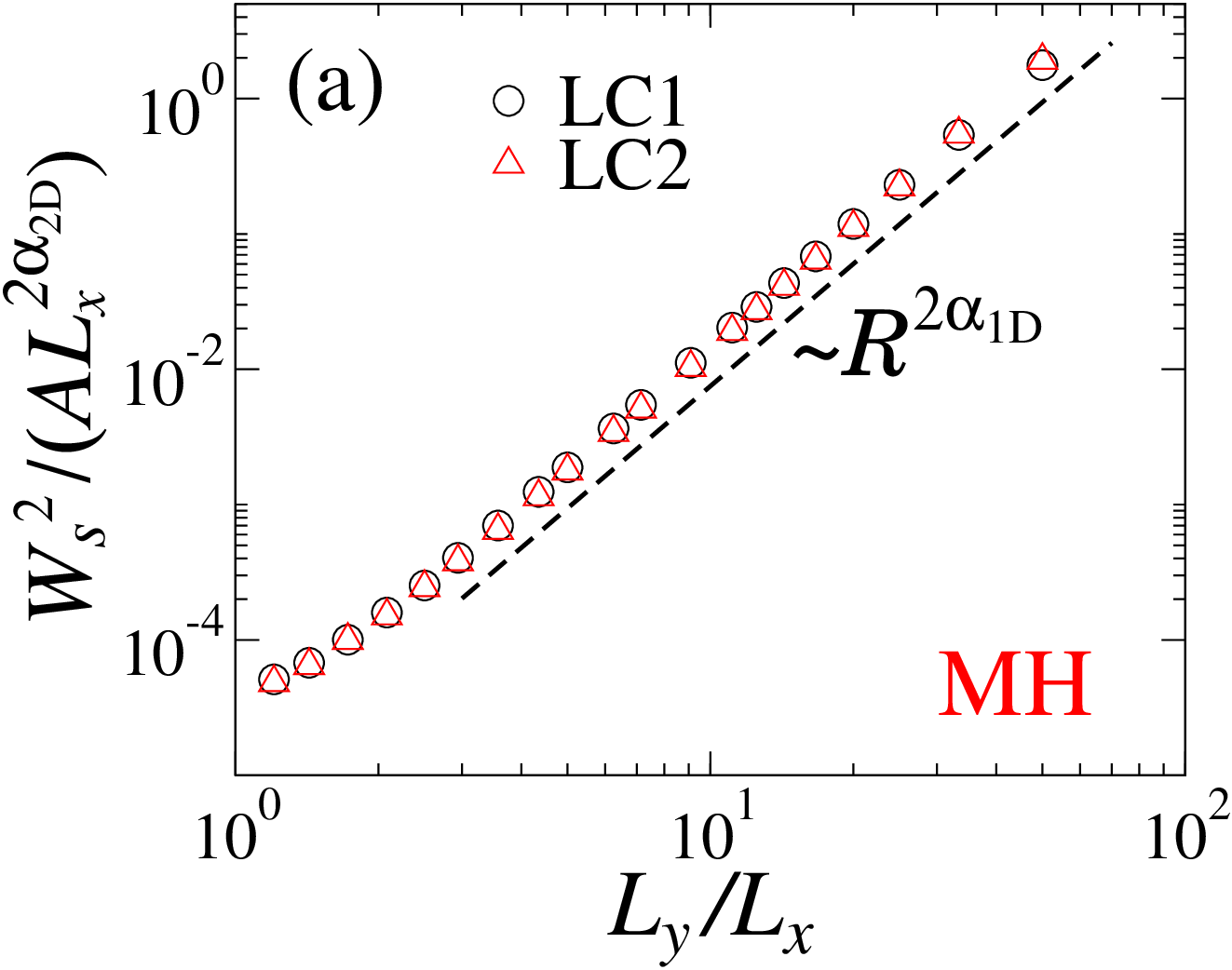}
 \includegraphics[width=4.2cm]{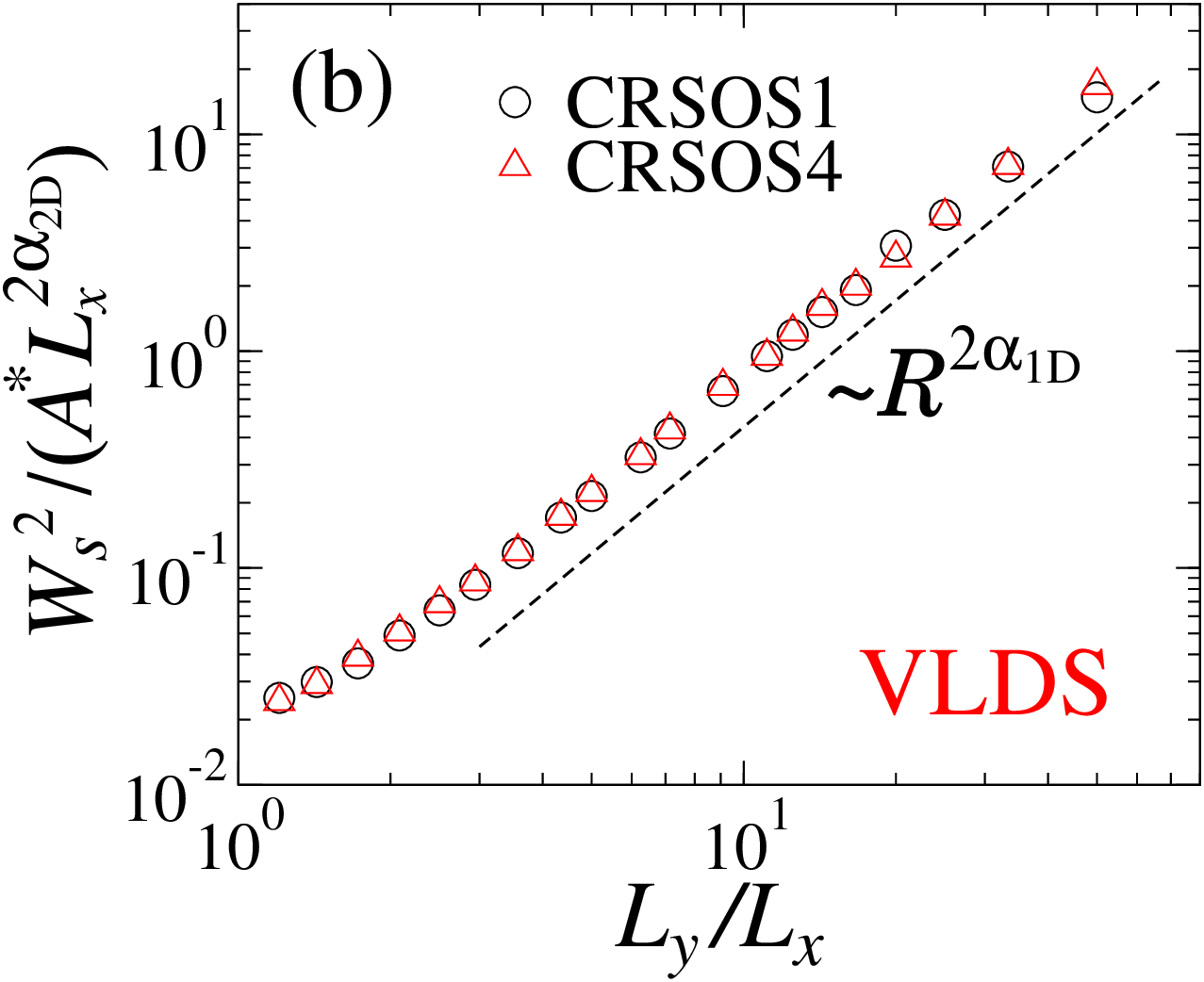}
 \includegraphics[width=4.2cm]{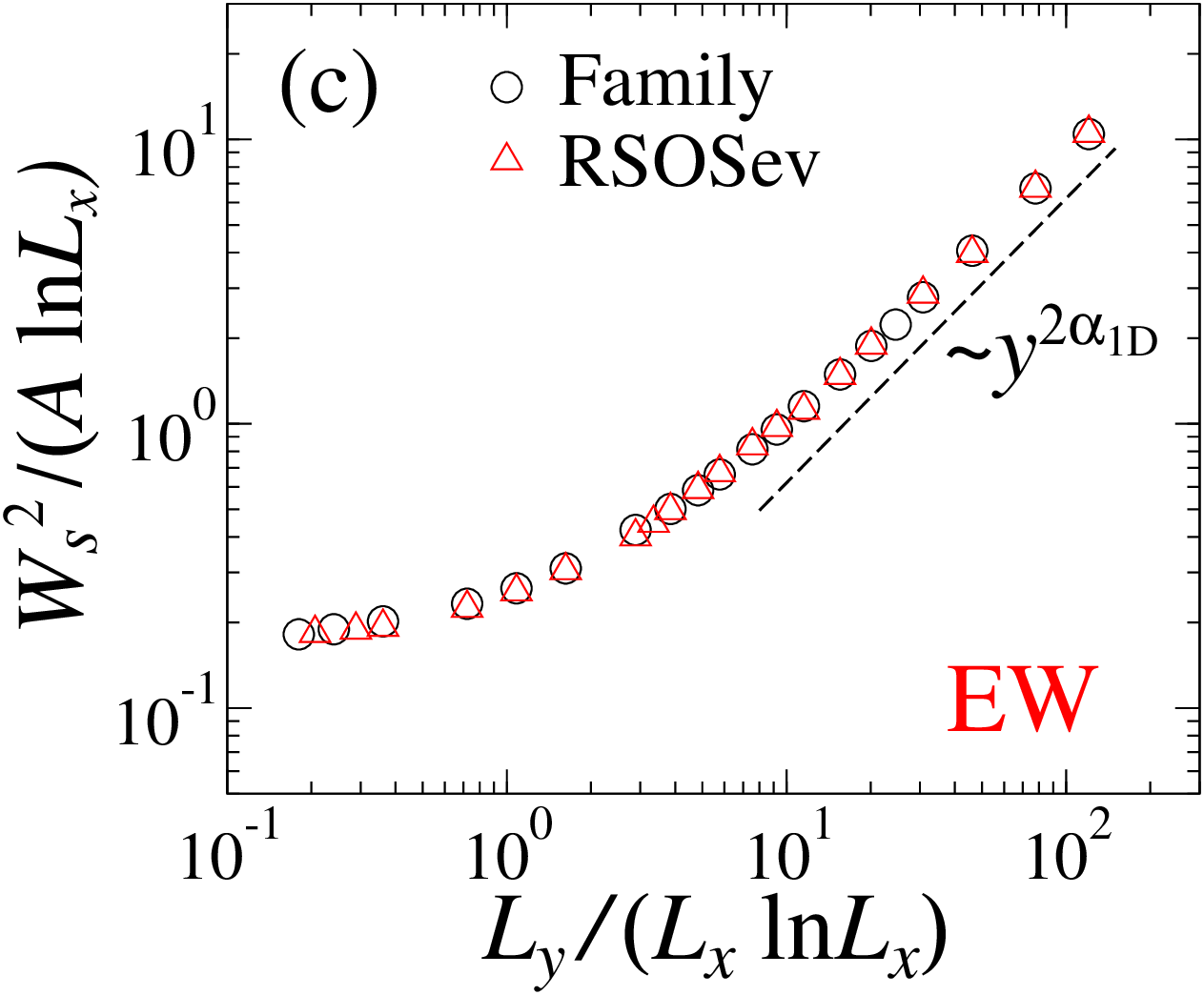}
 \includegraphics[width=4.2cm]{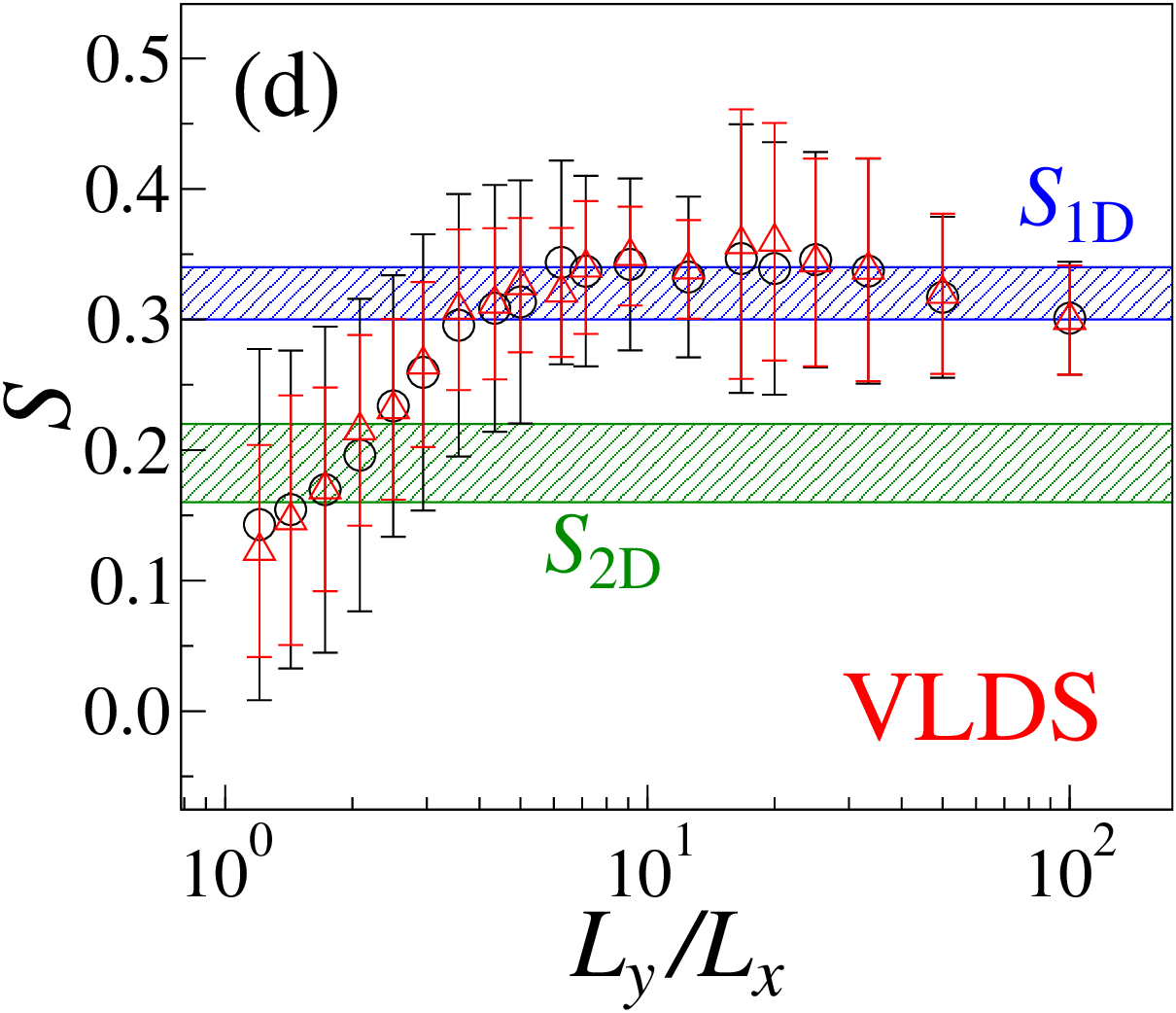}
  \caption{Rescaled curves of the squared steady state roughness $W_s^2$ versus the substrate aspect ratio $L_y/L_x$ [or $L_y/(L_x\ln L_x)$ in (c)] for the indicated models belonging to the (a) MH, (b) VLDS and (c) EW classes. The dashed lines have the indicated slopes. (d) Skewness $S$ against $L_y/L_x$ for the steady state HDs of the VLDS models. The shaded strips give the ranges of values estimated in the literature for such skewness in one- ($S_{1\text{D}}$) and two-dimensions ($S_{2\text{D}}$).}
\label{fig4}
\end{figure}

As previously established for the KPZ class in \cite{Ismael24}, the saturation roughness is expected to depend on the substrate aspect ratio, $\mathcal{R} = L_y/L_x$, according to the scaling relation in Eq.~\ref{eqScalingSS} or equivalently in Eq.~\ref{eqScalingSS2}. Indeed, applying this latter scaling to the MH and VLDS models, we find that it works very well to place all data --- for a given model and several substrate sizes --- into a single crossover curve, which increases asymptotically as $\mathcal{H}(\mathcal{R}) \sim \mathcal{R}^{\alpha_{1\text{D}}}$ [see Figs. \ref{fig4}(a) and \ref{fig4}(b)]. Moreover, as observed in Fig. \ref{fig4}(a), the curves for both LC models display a nice collapse when the values of $A$ are used to rescale them. Since $A$ is unknown for both CRSOS1 and CRSOS4 models, to verify the universality of their crossover curves, we have rescaled the CRSOS4 data by a constant factor $A^*=A_{\text{CRSOS4}}/A_{\text{CRSOS1}} = 16.34$. As shown in Fig. \ref{fig4}(b), this does indeed make it to collapse with the CRSOS1 curve (where we have considered $A^* =1$).

EW systems are, once again, expected to behave differently from the other classes, due to the logarithmic variation of the roughness in two-dimensions. Indeed, it suggests that the term $L_x^{2\alpha_{2\text{D}}}$ appearing in Eq. \ref{eqScalingSS2} should be replaced by $(\ln L_x)$ for the EW class. Moreover, when $L_y \gg L_x$, one has $W_s \sim \mathcal{R}^{\alpha_{1\text{D}}}$ for the other classes. Assuming that this also applies in the EW case, we arrive at the relation
\begin{equation}
 W_s^2 \simeq A (\ln L_x) \mathcal{H}_{EW}\left[ \frac{L_y}{L_x \ln L_x} \right],
 \label{eqScalingSSEW}
\end{equation}
where $\mathcal{H}_{EW}\left[y \right] \sim const.$ for $y \ll 1$ and $\mathcal{H}_{EW}\left[y \right] \sim y^{2\alpha_{1\text{D}}}$ for $y\gg 1$. Figure \ref{fig4}(c) shows the saturation roughness for the Family and RSOSev models rescaled according to Eq. \ref{eqScalingSSEW}. The very good collapse seen there confirms that this is indeed the crossover scaling for the EW class. As happened with the crossover time in the growth regime, the substrate aspect ratio has also acquired a multiplicative logarithmic `correction' in the EW class here. It is important to mention that, by applying the same reasoning above starting with Eq. \ref{eqScalingSS}, we obtain a scaling relation analogous to Eq. \ref{eqScalingSSEW}, but with both terms $(\ln L_x)$ replaced by  $\ln(L_x L_y)$. This alternative relation works so well as Eq. \ref{eqScalingSSEW} to collapse the EW data, because the logarithmic terms cancel out for large $\mathcal{R}$ and $\ln(L_x L_y) = \ln(L_x^2 \mathcal{R}) \sim \ln L_x$ when $\mathcal{R} \approx 1$.

\subsection{Height distributions}

Similarly to the growth regime, the HDs for the linear classes are Gaussian also in the steady state regime, in both one- and two-dimensions. Hence, no dimensional crossover is expected to be observed on them.

For the VLDS class, on the other hand, the steady state HDs are well-known to be different in one- and two-dimensions \cite{Reis2004VLDS,tiago07Rug}. For instance, the cumulant ratios $S_{\text{1D}}=0.32(2)$ and $K_{\text{1D}}=0.11(2)$ were reported for the 1D HD, whereas the values found for the 2D HD are $S_{\text{2D}}=0.19(3)$ and $K_{\text{2D}} \approx 0$ \cite{Reis2004VLDS,tiago07Rug}. Figure \ref{fig4}(d) presents the variation of $S$ with the aspect ratio $\mathcal{R}$ for both CRSOS models. Despite their large error bars, we clearly see that: \textit{i)} they collapse into a single curve, demonstrating that $\mathcal{R}$ is the only relevant quantity setting the HDs behavior; \textit{ii)} for small values of $\mathcal{R}$, they are consistent with previous estimates for $S_{\text{2D}}$, but then they converge to $S_{\text{1D}}$ as $\mathcal{R}$ increases. The slightly smaller values obtained here, in comparison with the central ones in the literature, are likely due to the absence of finite-size extrapolations. (Note that, in order to do this, we should have to simulate several sizes, for a given $\mathcal{R}$, and then to repeat this procedure for various aspect ratios.) It is curious that the skewness attains the 1D value already for $\mathcal{R} \approx 5$, indicating that not so large aspect ratios are needed to observe HDs similar to the 1D HD. This is different from the KPZ scenario, where $S$ (and $K$, as well) varies slowly between the 2D and 1D values as $\mathcal{R}$ augments.

\section{The $L_x = L_y^{\delta}$ case}
\label{secPG}

While the scaling relations in Eqs. \ref{eqCrossScaling} and \ref{eqScalingSS} (or, equivalently, \ref{eqScalingSS2}) are valid for general $L_x$ and $L_y$, it is interesting to analyze the particular situation where $L_x = L_y^{\delta}$, with an exponent $\delta \in [0,1]$, such that the usual 1D and 2D systems are recovered for $\delta=0$ and $\delta=1$, respectively.

\subsection{Growth regime}

Let us start discussing the implications of this for the temporal crossover, considering that $L_y$ (and then $L_x$) is kept fixed during the growth process. Firstly, we recall that the 2D-to-1D crossover occurs at a time $t_c \sim L_x^{z_{2\text{D}}} \sim L_y^{\delta z_{2\text{D}}}$ [or $t_c \sim (L_x \ln L_x)^{z_{2\text{D}}} \sim  (\delta L_y^{\delta} \ln L_y)^{2}$ in the EW case]. Later on, in the saturation time $t_s$, the surface becomes fully correlated also in the $y$ direction, attaining thus the steady state regime. Therefore, in order to observe the dimensional crossover, one must have $t_s \gg t_c$; otherwise, the onset of the 1D scaling will be preempted by saturation. As demonstrated in Ref. \cite{Ismael24} for KPZ and observed here for the other classes, when the aspect ratio $\mathcal{R}$ is not so large, one has $t_s \sim L_y^{z_{2\text{D}}}$. This yields $t_s/t_c \sim L_y^{z_{2\text{D}}(1-\delta)}$, which diverges for any $0 < \delta < 1$ in the $L_y \to \infty$ limit. It turns out, however, that $t_s \sim L_y^{z_{1\text{D}}}$ when $L_y \gg L_x$ --- i.e., the dynamic exponent changes to the 1D one --- and, then, in this asymptotic situation
\begin{equation}
t_s/t_c \sim L_y^{z_{1\text{D}} - \delta z_{2\text{D}}},
\label{eqtstc}
\end{equation}
which only diverges, as $L_y \to \infty$, if $\delta < z_{1\text{D}}/z_{2\text{D}}$. Hence, there exists a `special' exponent
\begin{equation}
\delta^* = \frac{z_{1\text{D}}}{z_{2\text{D}}}
\end{equation}
above which the 2D-to-1D crossover shall not be observed. Namely, for $\delta \geq \delta^*$, $L_y$ will never become sufficiently larger than $L_x$ to yield $t_s \gg t_c$ and, consequently, the dimensional crossover may not show up.

\begin{figure}[t]
 \includegraphics[width=4.2cm]{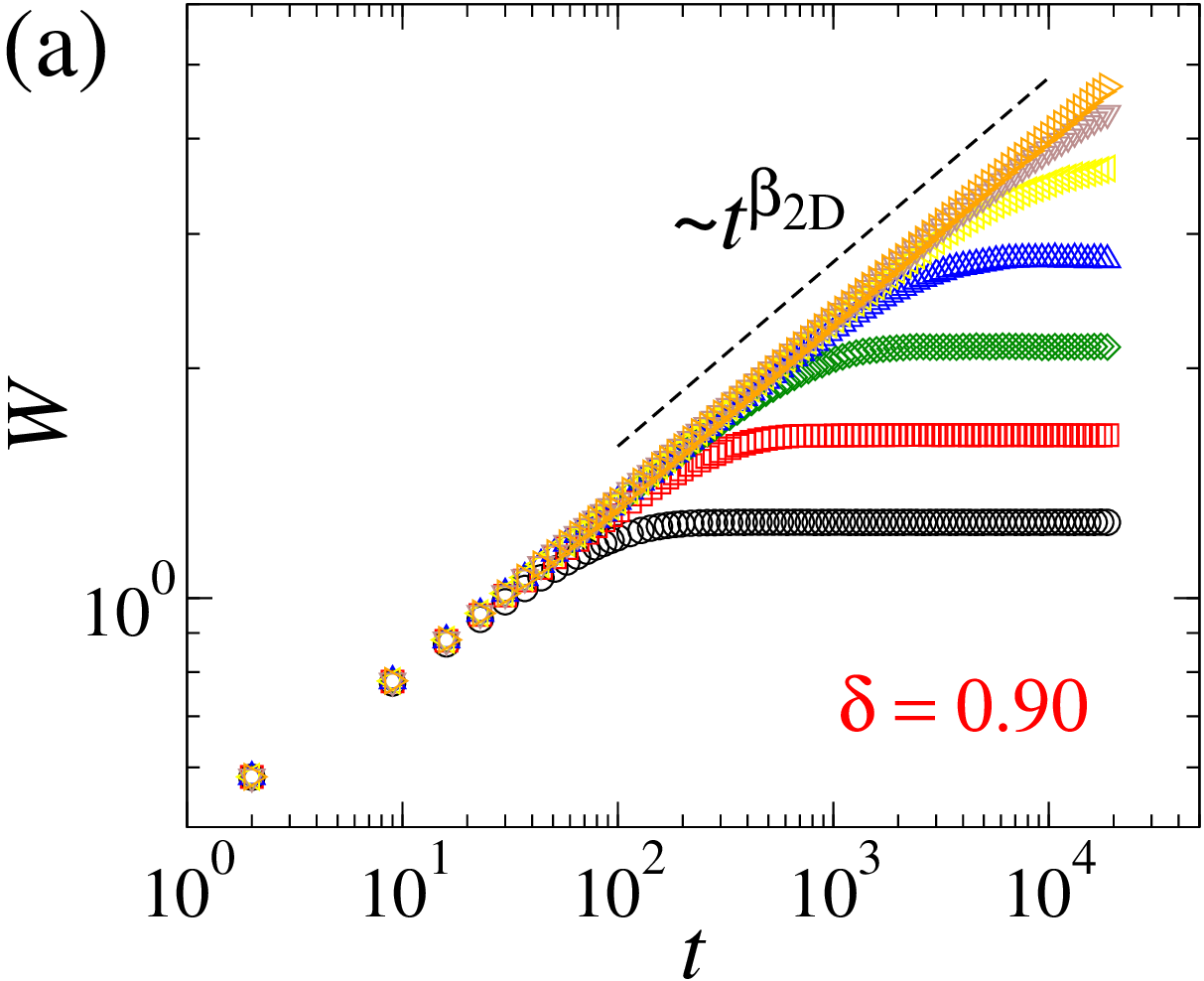}
 \includegraphics[width=4.2cm]{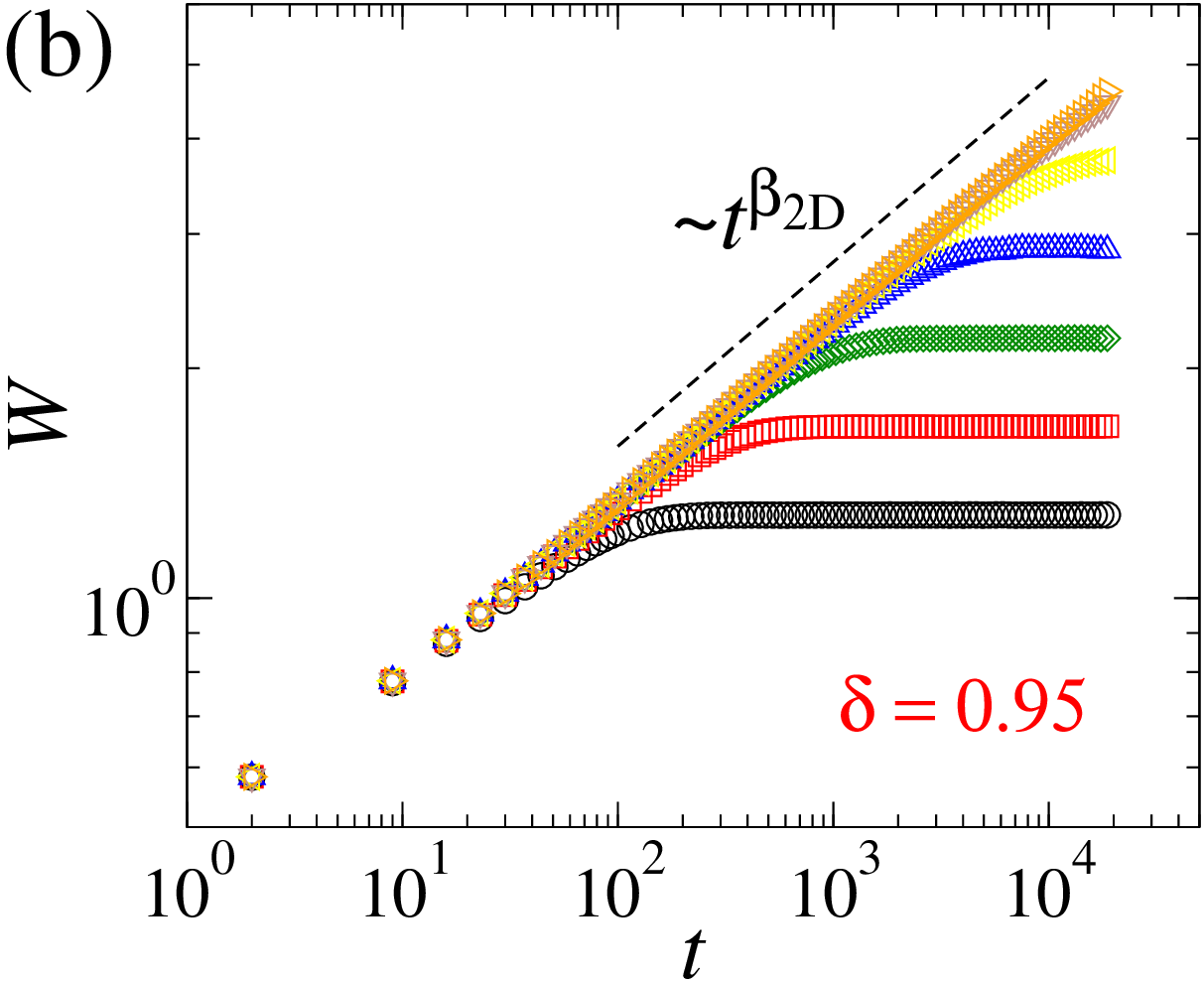}
 \caption{Temporal evolution of the surface roughness for the RSOS model deposited on substrates with $L_x = L_y^{\delta}$ for (a) $\delta<\delta_{KPZ}^*$ and (b) $\delta> \delta_{KPZ}^*$. In both panels, data for $L_y = 32, 64, \ldots, 2048$ are shown. The dashed lines have the indicated slopes.}
\label{figPG1}
\end{figure}

We remark that this scenario cannot happen for the linear (EW and MH) classes, because their dynamic exponents are independent of the dimensionality, so that $\delta^*=1$ for them. Although Eq. \ref{eqtstc} changes to $t_s/t_c \sim [L_y^{1 - \delta}/\ln L_y]^2$ in the EW case, this conclusion is still the same. On the other hand, the `special' behavior above can play an important role for the nonlinear (KPZ and VLDS) classes. For instance, in the KPZ case, $z_{1\text{D}}=3/2$ and $z_{2\text{D}} \approx 1.611$ \cite{tiago22} give $\delta_{KPZ}^* \approx 0.931$. For the VLDS class, the 2-loop exponents $z_{1\text{D}} \approx 2.939$ and $z_{2\text{D}} \approx 3.306$ yield $\delta_{VLDS}^* \approx 0.889$.

However, since the effect of $\delta > \delta^*$ is only expected in the asymptotic ($L_y \rightarrow \infty$) limit, it is very hard to confirm its existence numerically. For instance, almost identical results are found in Figs. \ref{figPG1}(a) and \ref{figPG1}(b) in the temporal variation of $W$ for the RSOS model, considering values of $\delta$ close to $\delta_{KPZ}^*$, with one value below and the other above the `special' exponent. This indicates that much larger sizes $L_y$ and much longer deposition times are needed to observe the 2D-to-1D crossover for $\delta = 0.90$ (and its absence for $\delta = 0.95$). We have verified, for example, that even data [not shown] for $L_y=8192$ up to $t=10^5$ do not present any noticeable difference for these $\delta$'s. Also, we have found a similar situation for the VLDS class, where simulations with $\delta=0.85$ and $\delta=0.95$, for example, yield almost identical results [data not shown].

\subsection{Steady state regime}

\begin{figure}[!t]
 \includegraphics[width=4.2cm]{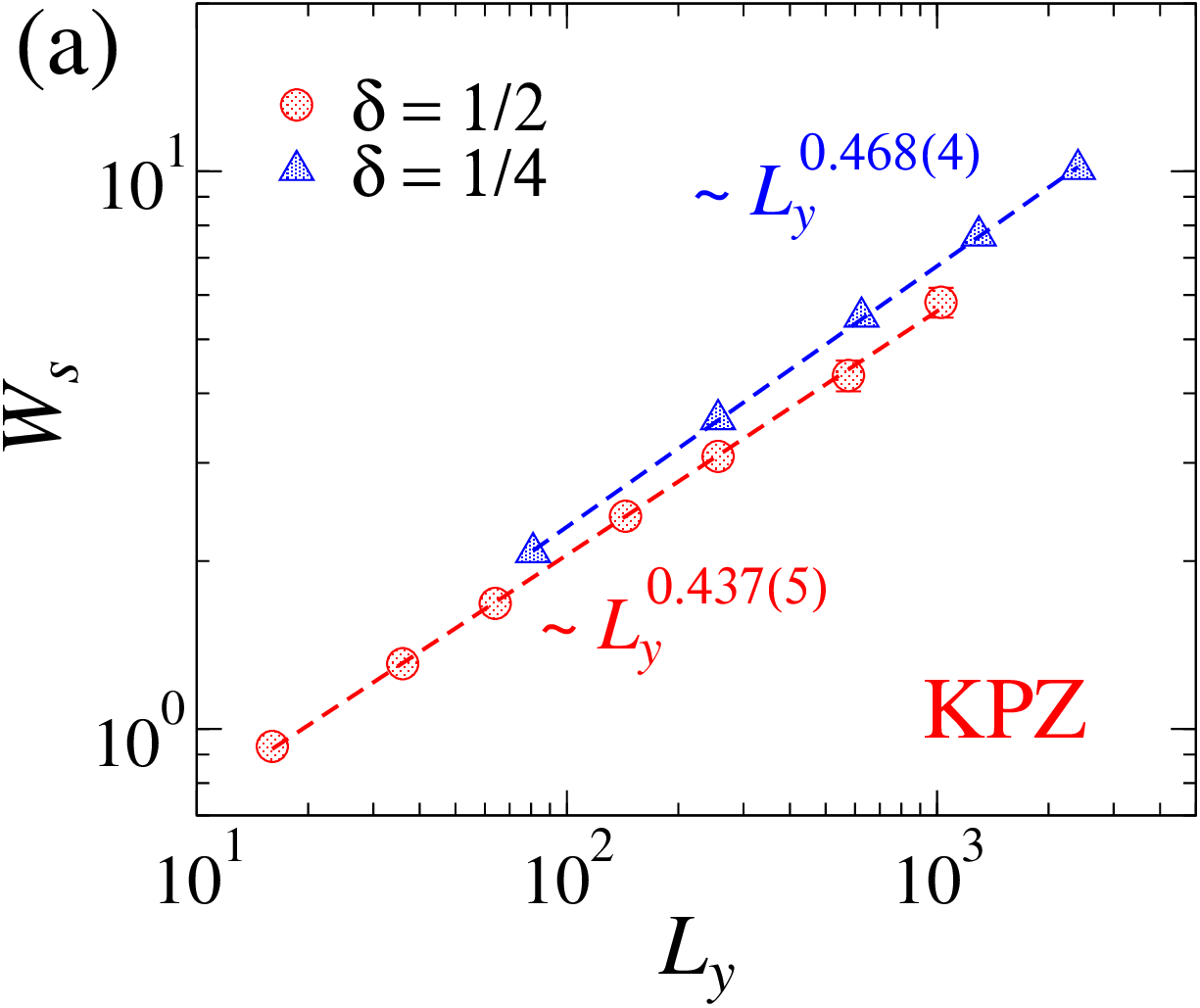}
 \includegraphics[width=4.2cm]{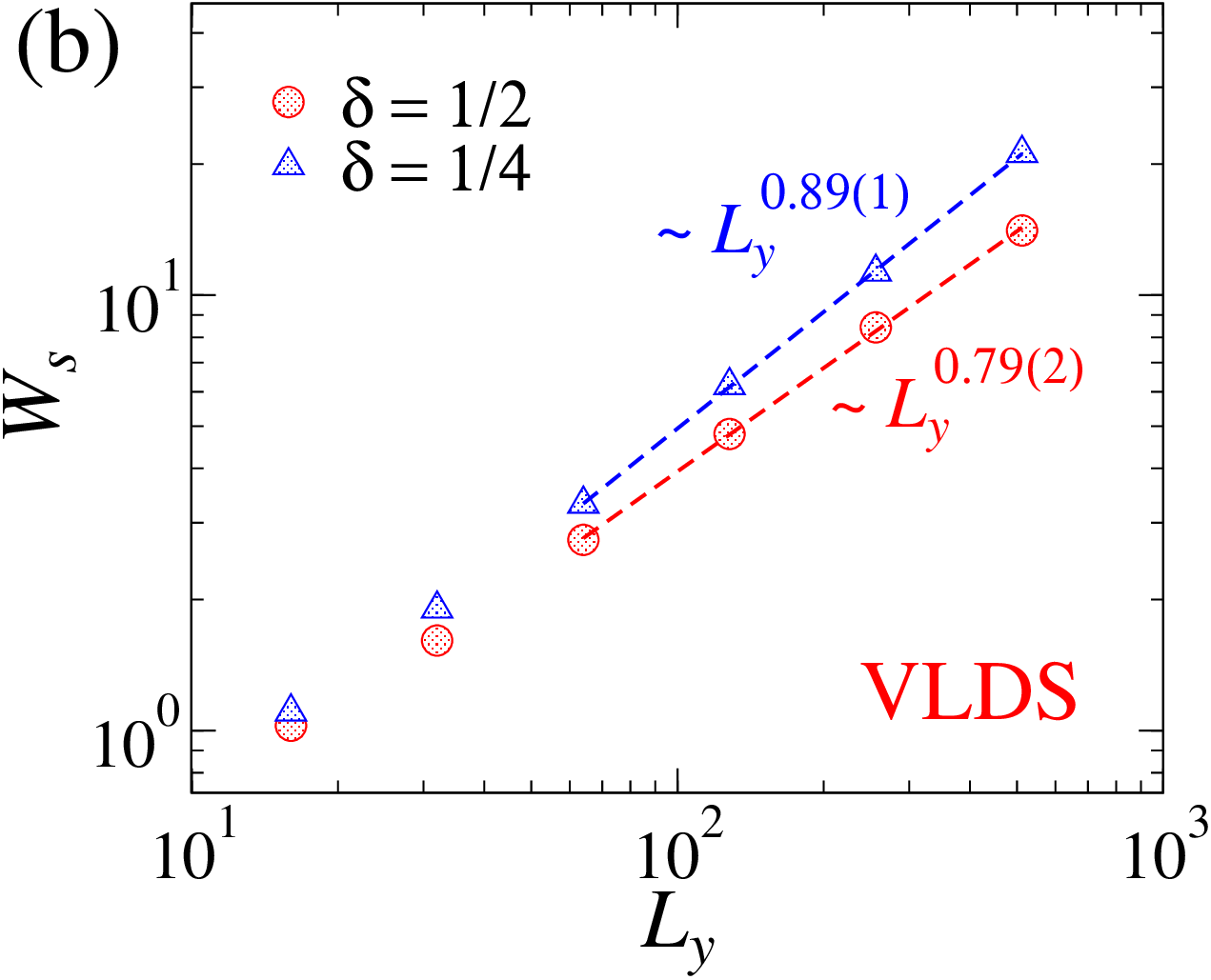}
 \includegraphics[width=4.2cm]{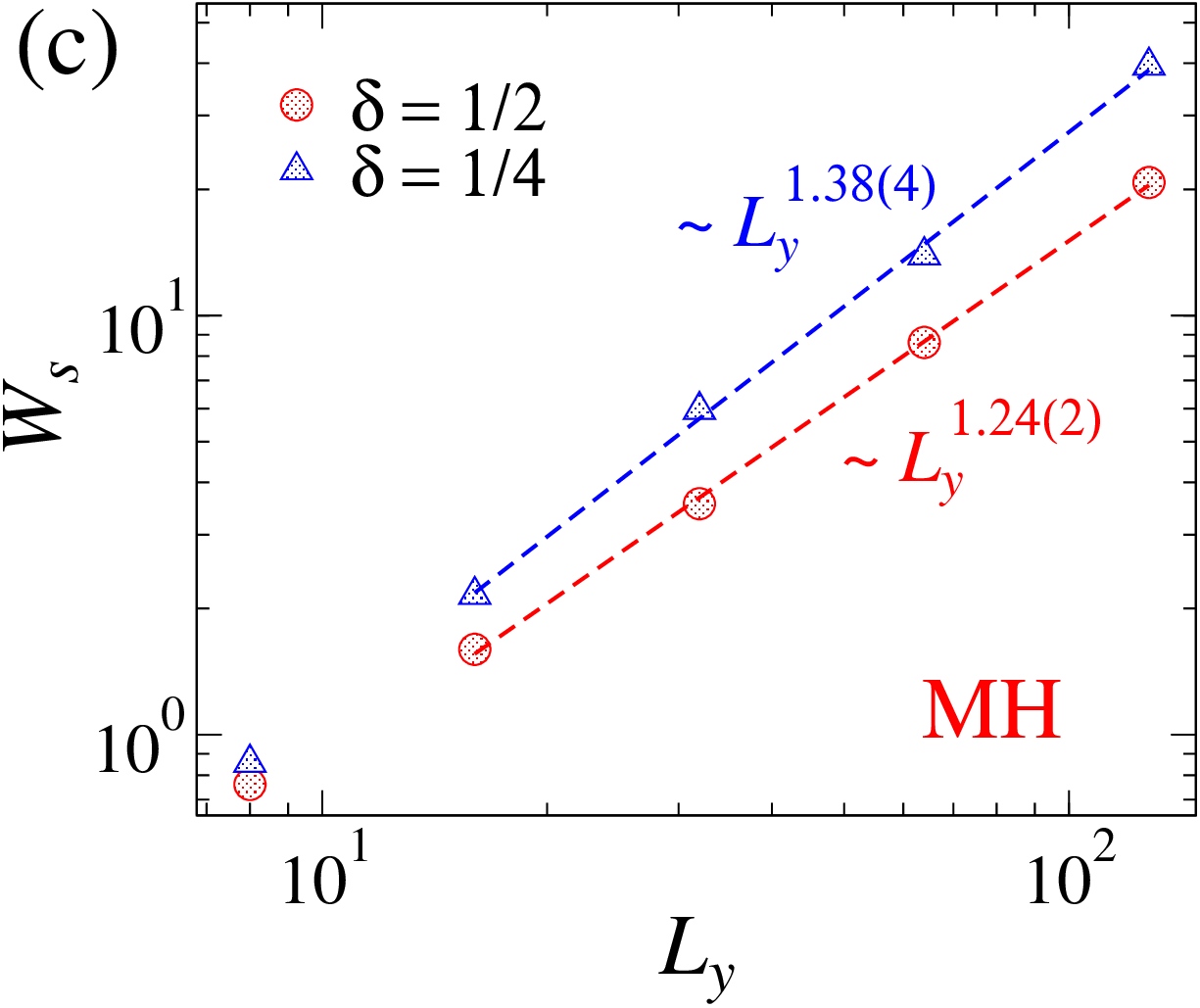}
 \includegraphics[width=4.2cm]{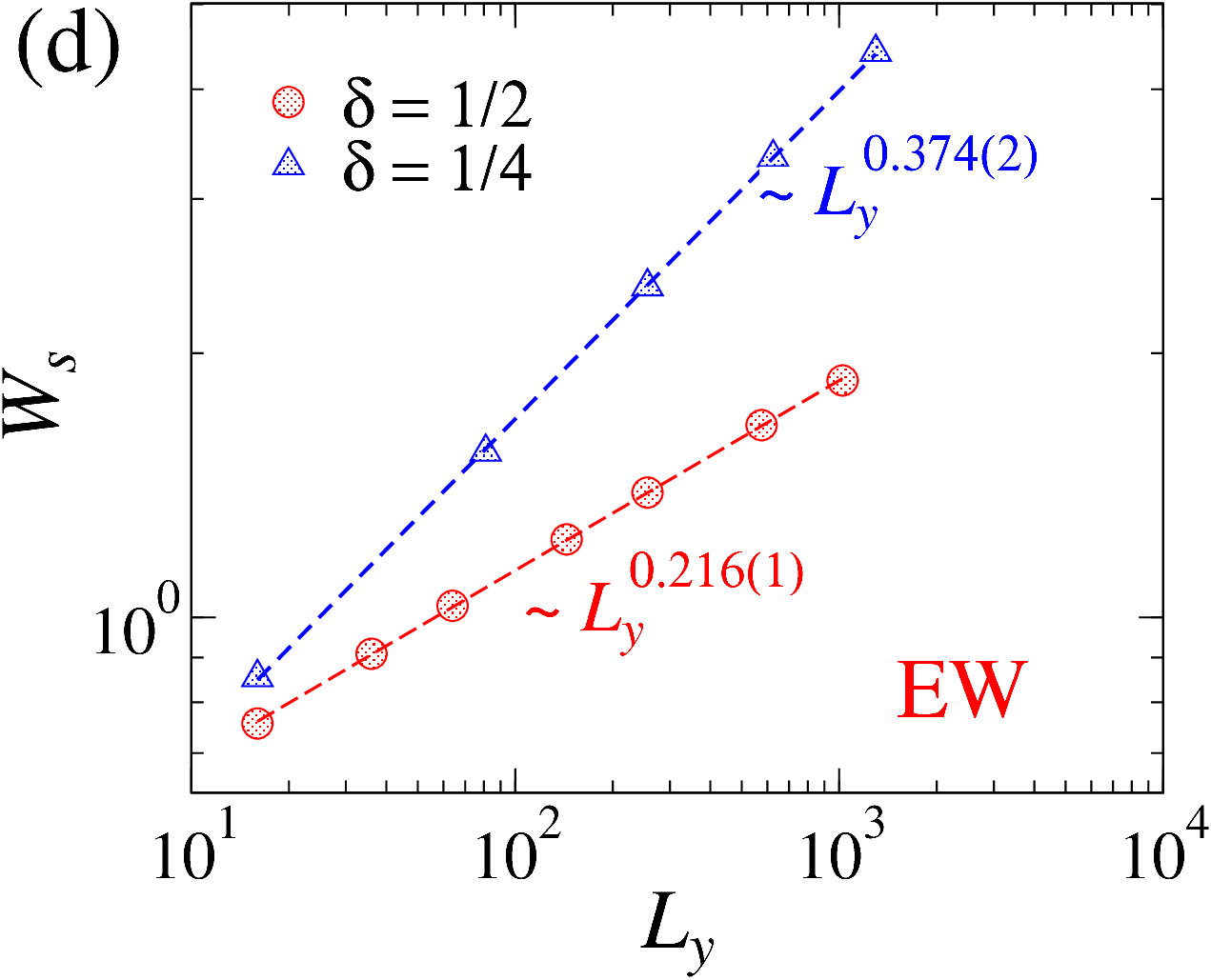}
 \caption{Saturation roughness $W_s$ versus $L_y$ for the (a) RSOS, (b) RSOSev (c) LC1 and (d) CRSOS1 models. Results for different exponents $\delta$ are shown in each case. The dashed lines are linear fits of the data and have the indicated slopes. The error bars in such slopes were obtained by fitting different regions within the ranges considered.}
\label{figPG2}
\end{figure}

Next, we focus on the saturation regime. By replacing $L_x$ by $L_y^{\delta}$ in the scaling for the saturation roughness (Eq. \ref{eqScalingSS2}), one gets
\begin{equation}
W_s^2 \sim L_y^{2\delta\alpha_{2\text{D}}} \mathcal{H}( L_y^{1 - \delta} ),
\end{equation}
where we recall that $\mathcal{H}(\mathcal{R}) = const.$ for $\mathcal{R} = 1$ (i.e., for $\delta=1$) and $\mathcal{G}(\mathcal{R}) \sim \mathcal{R}^{2 \alpha_{1\text{D}}}$ for $\mathcal{R} \gg 1$. This means that, in this last regime,
\begin{equation}
W_s \sim L_y^{\Lambda},
\end{equation}
with
\begin{equation}
\Lambda = \delta \alpha_{2\text{D}} + (1 - \delta) \alpha_{1\text{D}}.
\label{eqLamb}
\end{equation}
Thereby, the 1D scaling [$W_s \sim L_y^{\alpha_{1\text{D}}}$] is recovered for $\delta = 0$ and, exactly at $\mathcal{R} = 1$, it gives the correct 2D behavior [$W_s \sim L_y^{\alpha_{2\text{D}}}$], even though we have no guarantee that this expression for $\Lambda$ is valid for $\mathcal{R} \approx 1$. Note also that the condition $\mathcal{R} = L_y^{1-\delta} \gg 1$ is satisfied for any $\delta<1$ provided that $L_y$ is large enough. Hence, Eq. \ref{eqLamb} is valid for general $\delta \in [0,1]$ in the asymptotic $L_y \rightarrow \infty$ limit. In practice, we may see in Fig. \ref{fig4} an increase consistent with $\mathcal{G}(\mathcal{R}) \sim \mathcal{R}^{2 \alpha_{1\text{D}}}$ appearing already for $\mathcal{R} \geq \mathcal{R}_{min} \approx 3$ for the MH and VLDS classes; and a similar result (but with $\mathcal{R}_{min} \approx 5$) was found in Fig. 3(c) of Ref. \cite{Ismael24} for the KPZ models. This suggests that Eq. \ref{eqLamb} will hold for $\delta \leq 1- \frac{\ln \mathcal{R}_{min}}{\ln L_y}$, for finite $L_y$. In any case, it is quite interesting that the `roughness exponent' $\Lambda$ is a mixture of the 1D and 2D ones, weighted by $\delta$.

This behavior is confirmed in Figs. \ref{figPG2}(a)-(c) for the KPZ, VLDS and MH classes. We notice that in the KPZ case, $\Lambda$ is limited to the range $0.388 \lesssim \Lambda \leqslant 0.5$, so that no large variation is expected with $\delta$. For example, for $\delta = 1/2$ and $\delta = 1/4$, one has $\Lambda \approx 0.44$ and $\Lambda \approx 0.47$, respectively, which are very close to the values found in Fig. \ref{figPG2}(a) from simple power-law fits to the data.

For the VLDS and MH classes, we have $\Lambda_{VLDS} \in [0.653,0.969]$ and $\Lambda_{MH} \in [1,3/2]$. Hence, more clear differences can be observed in the scaling of $W_s$ for them, depending on $\delta$, as we may indeed see in Figs. \ref{figPG2}(b) and \ref{figPG2}(c). For instance, in the VLDS case, we expect to have $\Lambda = 0.811$ for $\delta=1/2$ and $\Lambda = 0.890$ for $\delta=1/4$. The exponent found in Fig. \ref{figPG2}(b) for $\delta=1/4$ is fully consistent with this, while a slightly smaller (though very close) result is obtained for $\delta=1/2$. This discrepancy is likely due to finite-size effects, as we get even smaller exponents when the first two discarded points are included in the fits. For the MH class, considering again $\delta=1/4$ and $1/2$, the exponents predicted by Eq. \ref{eqLamb} are $\Lambda = 1.375$ and $\Lambda = 1.25$, respectively, which are confirmed by the numerical results displayed in Fig. \ref{figPG2}(c) for the LC1 model.

For EW systems, we may expect that $\Lambda_{EW} \in [0,1/2]$ and, since $\alpha_{2\text{D}}=0$, equation \ref{eqLamb} should simplify to $\Lambda_{EW} = (1-\delta) \alpha_{1\text{D}}=(1-\delta)/2$. This is indeed the case when $\delta$ is small, so that $L_y \gg L_x$. As a matter of fact, for $\delta=1/4$ we expect that $\Lambda_{EW} = 3/8$ and the power-law fit in Fig. \ref{figPG2}(d) returns $\Lambda_{EW} = 0.374(2)$. In contrast, the numerical results in Fig. \ref{figPG2}(d) for $\delta=1/2$ yields $\Lambda_{EW} = 0.216(1)$, which is a bit smaller than $1/4$. Recalling that we must have a logarithmic behavior for the EW roughness when $\delta=1$, it is not a surprise that Eq. \ref{eqLamb} will begin to fail for this class at large $\delta$.

\section{Summary}
\label{secConc}

We have investigated the dimensional crossover in the kinetic roughening of self-affine surfaces growing on rectangular substrates, for several universality classes. When $L_y \gg L_x$, we found that all systems behave as if they were 2D at short times ($t \ll t_c$), with the roughness scaling in time with the exponent $\beta_{2\text{D}}$. This corresponds to a logarithmic variation in the EW class and, consequently, the general crossover scaling obtained for the other classes [Eq. \ref{eqCrossScaling}] had to be modified in the EW case. For instance, the crossover time, which is $t_c \sim L_x^{z_{2\text{D}}}$ in the KPZ, VLDS and MH classes, changes to $t_c \sim [L_x \ln L_x]^{z_{2\text{D}}}$ for EW systems. At $t \sim t_c$ [when the lateral correlations reach the smaller size $L_x$], the growth dynamics cross over to a 1D regime, where the roughness increases as $t^{\beta_{1\text{D}}}$.

While in the KPZ class this crossover in the roughness is accompanied by a corresponding 2D-to-1D crossover in the height distributions (HDs), such behavior is less evident for the other cases. In fact, in the MH and EW classes, the HDs are Gaussian in both dimensions and no appreciable difference was seen in their cumulant ratios between the 2D and 1D regimes. The VLDS HDs are non-Gaussian and display a crossover in the convergence of the skewness and kurtosis, although their asymptotic values are nearly identical for the 1D and 2D HDs.

At a time $t_s$, when the correlations have spread over the entire surface, the roughness saturates. In this steady-state regime, we found that the saturation roughness for all classes follow the same scaling with the substrate aspect ratio [given by Eqs.~\ref{eqScalingSS} or \ref{eqScalingSS2}], but it had also to be adapted for EW systems, in order to account for their logarithmic behavior with the substrate size in two-dimensions. Moreover, for the VLDS class, the skewness of the steady-state HDs clearly exhibits a crossover from the 2D to the 1D value as the substrate aspect ratio increases. This indicates that there exists a continuous family of distributions, interpolating between the 2D and 1D ones. These results demonstrate that the dimensional crossover extends beyond the roughness scaling and likely affects all universal properties of the growing surfaces.

By rewriting the system size as $L_x = L_y^{\delta}$, where $\delta \in [0,1]$, we have also identified a `special' exponent $\delta^*$ above which the characteristic times $t_c$ and $t_s$ are of the same order, hindering thus the appearance of the 1D regime in the system's evolution. This `critical' value is given by $\delta^* = z_{1\text{D}} / z_{2\text{D}}$, yielding $\delta^*_{\text{KPZ}} \approx 0.931$ and $\delta^*_{\text{VLDS}} \approx 0.889$. For the linear classes, one has $\delta^* = 1$, meaning that the dimensional crossover only disappear for square substrates (i.e., for $\delta=1$). Unfortunately, confirming these interesting predictions numerically would require  simulations for system sizes and deposition times far beyond our current computational capabilities.

Another key result revealed by the $L_x = L_y^{\delta}$ condition is the fact that the saturation roughness presents a non-universal scaling with the system size, i.e., $W_s \sim L_y^{\Lambda}$ with a $\delta$-dependent exponent $\Lambda$. We recall that claims about a possible breakdown of universality in growing systems have appeared several times in the literature, but it seems that most of them were refuted in subsequent works. Here, we have a genuine situation where such breakdown occurs, as a consequence of the nonequilateral substrate sizes. Interestingly, while the steady state regime presents this nonuniversality, in the growth regime the dynamics is universal, in the sense that it always selects between the 1D or 2D behavior.

From a theoretical standpoint, our results provide strong evidence that the dimensional crossover originally identified for KPZ systems is a general feature in surface growth phenomena and should be expected across all universality classes. Therefore, it may appears also in real systems, provided that $L_x$ is small enough to make the crossover time $t_c$ accessible in the experimental times, as it may be the case, e.g., in the growth of rectangular nanostructures.

\acknowledgments

The authors are grateful to Peter Grassberger for insightful discussions and for a critical reading of the manuscript. They also acknowledge partial financial support from the Brazilian agencies CNPq, FAPEMIG, and FAPDF (grant number 00193-00001817/2023-43).

\appendix

\section{Non-universal parameters for the 2D RSOSev model}

\begin{figure}[b]
 \includegraphics[width=4.1cm]{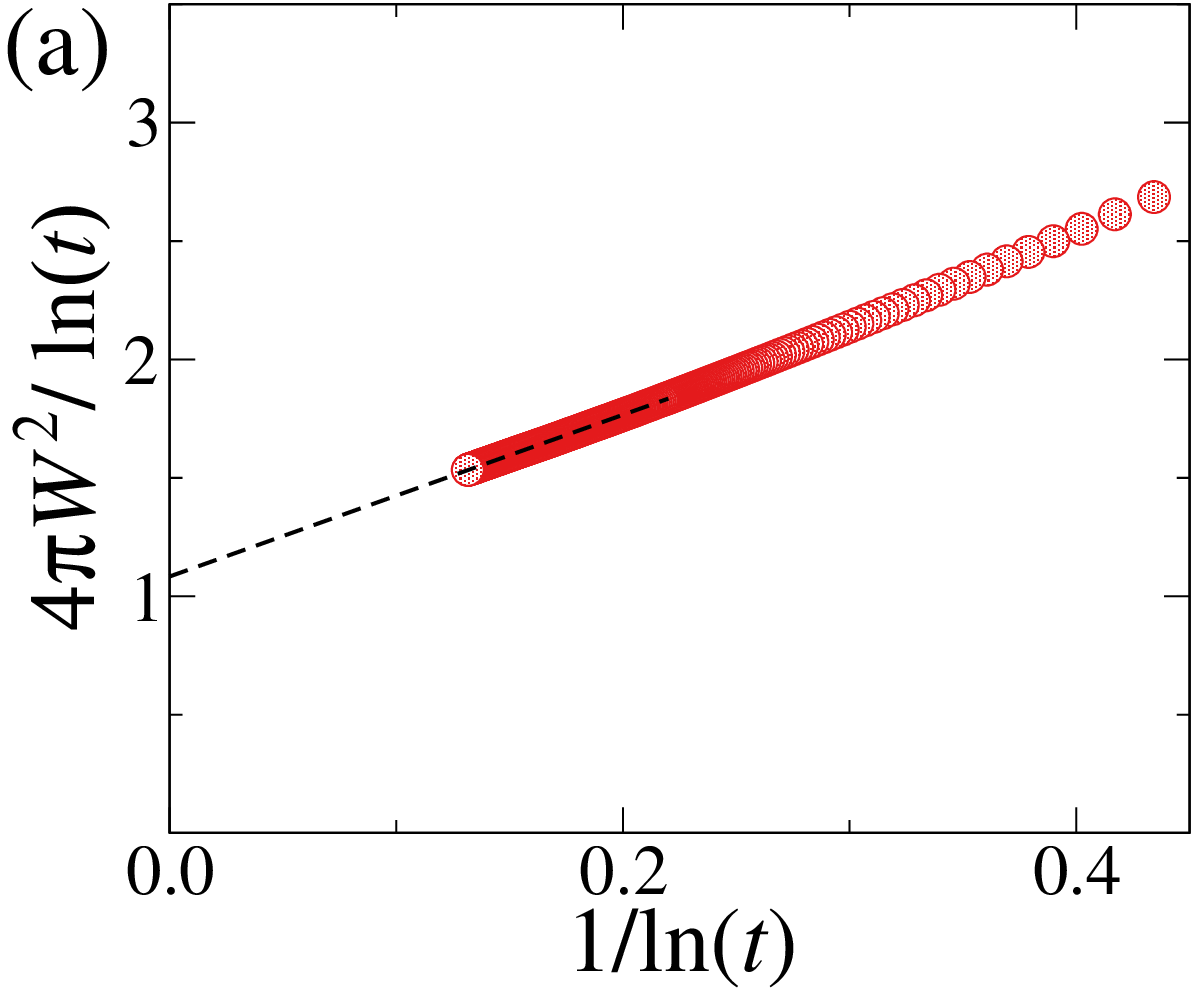}
 \includegraphics[width=4.3cm]{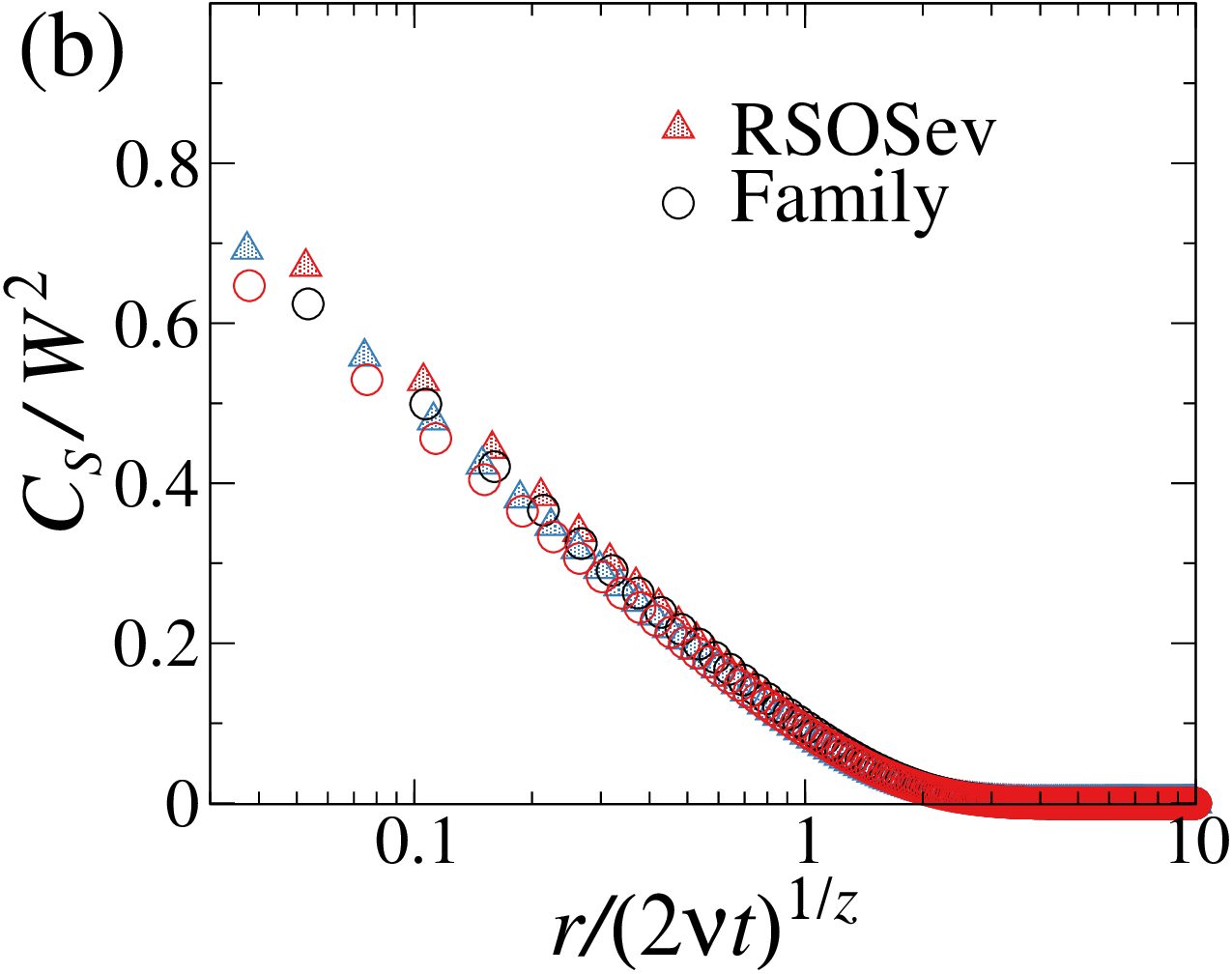}
\caption{(a) Rescaled squared roughness $W^2/\ln t$ versus $1/\ln t$ for the 2D RSOSev model. The dashed line is a linear fit to the data used to extrapolate it to the $t \rightarrow \infty$ limit. (b) Rescaled covariance $C_S/W^2$ as function of the rescaled length $r/(2\nu t)^{1/z}$ (with $z=2$) for the 2D Family and RSOSev models, where curves for times in the range $250 \leqslant t \leqslant 2000$ are shown. The results in both plots are from simulations of these models on square substrates of sizes $L_x=L_y=2048$.}
\label{figAP}
\end{figure}

We will obtain here the non-universal parameters $A$ and $\nu$ for the 2D RSOSev model using the same procedures employed in Ref. \cite{Ismael19} for other EW models. As discussed there, the global roughness of 2D EW systems increases in the growth regime as $W^2 \simeq A \langle \chi^2\rangle_c \ln t$, where $\langle \chi^2\rangle_c = 1/4 \pi$ is the universal variance of the underlying HD in this regime \cite{Ismael19}. Thereby, the non-universal amplitude $A$ can be obtained by extrapolating $4\pi W^2/\ln t$ to the $t \rightarrow\infty$ limit. Similarly to other 2D EW models \cite{Ismael19}, a good linear behavior is observed is such a extrapolation by using $1/\ln t$ in the abscissa [see Fig. \ref{figAP}(a)]. Linear fits considering different intervals of this curve, yields $A = 1.06(6)$.

In order to estimate the coefficient $\nu$, we may use the 2-point spatial covariance: $C_S(r,t) = \langle \tilde{h}(\vec{x}+\vec{r},t)\tilde{h}(\vec{x},t) \rangle$, with $\tilde{h}(\vec{x},t) \equiv h(\vec{x},t)-\overline{h}(t)$. Since $C_S(r=0,t) = W^2(t)$, rescaled curves of $C_S/W^2$ versus $r/\xi$ are expected to collapse for different times and models, if one uses the correlation length parallel to the surface, $\xi \simeq (2\nu t)^{1/z}$, for EW systems, with the correct $\nu$. Figure \ref{figAP}(b) shows covariance curves rescaled in this way for the 2D Family model, for which we already known, from Ref. \cite{Ismael19}, that $\nu \approx 0.70$. Then, by rescaling the data for the 2D RSOSev model in the same way, we may find the value of $\nu$ that yields the best collapse with the Family curves. This procedure gives $\nu = 0.18(4)$ for 2D RSOSev model. We notice that the very good collapse of the curves of roughness displayed in Fig. \ref{fig1}(f), where this value was used to rescale the data, provides additional evidence that it is correct.

\bibliography{bibExpKPZ2D}

\end{document}